%% file: paper.tex
\documentclass{article}

\usepackage[bookmarks,bookmarksopen,bookmarksdepth=2]{hyperref}
\hypersetup{pdfpagemode=UseNone}

\usepackage{xcolor}
\usepackage{graphicx}
\usepackage{caption}
\usepackage{subfig}
\usepackage{ifthen}

\usepackage{amssymb}
\usepackage[numbers]{natbib}

\usepackage{tabularx}
\usepackage{multirow}
\usepackage{booktabs}

\usepackage{amsthm}
\theoremstyle{definition}
\newtheorem{defn}{Definition}

\newcommand{\eg}{e.g.,~}							
\newcommand{\ie}{i.e.,~}							
\newcommand{\etal}{~et al.}					
\newcommand{\Fig}[1]{Figure~\ref{#1}}  			
\newcommand{\Table}[1]{Table~\ref{#1}}	    
\newcommand{\Sect}[1]{Section~\ref{#1}}	  
\newcommand{\Code}[1]{\texttt{\small{#1}}}	

\usepackage{authblk}

\begin{document}

\title{Systematic Mapping Study of Template-based Code Generation}


\author[1]{Eugene Syriani\thanks{syriani@iro.umontreal.ca}}
\author[1]{Lechanceux Luhunu\thanks{luhunukl@iro.umontreal.ca}}
\author[1]{Houari Sahraoui\thanks{sahraoui@iro.umontreal.ca}}
\affil[1]{University of Montreal, Canada}
\renewcommand\Authands{ and }

\maketitle

\begin{abstract}
Template-based code generation (TBCG) is a synthesis technique that produces code from high-level specifications, called templates. TBCG is a popular technique in model-driven engineering (MDE) given that they both emphasize abstraction and automation. Given the diversity of tools and approaches, it is necessary to classify existing TBCG techniques to better guide developers in their choices.
The goal of this article is to better understand the characteristics of TBCG techniques and associated tools, identify research trends, and assess the importance of the role of MDE in this code synthesis approach.
We conducted a systematic mapping study of the literature to paint an interesting picture about the trends and uses of TBCG.
Our study shows that the community has been diversely using TBCG over the past 15 years. TBCG has greatly benefited from MDE. It has favored a template style that is output-based and high level modeling languages as input. TBCG is mainly used to generate source code and has been applied in a variety of domains. Furthermore, both MDE and non-MDE tools are becoming effective development resources in industry.

\end{abstract}

\input{introduction}

\input{relatedwork}

\input{methods}

\input{selection}

\input{trends}

\input{characteristics}

\input{correlations}

\input{tools}

\input{mde}

\input{discussion}

\input{conclusion}


\bibliographystyle{spmpsci}      
\bibliography{bibliography,final}   

\end{document}

%% file: introduction.tex
\section{Introduction} \label{sec:introduction}

Code generation has been around since the 1950s, taking its origin in early compilers~\cite{Rich1988}.
Since then, software organizations have been relying on code synthesis techniques in order to reduce development time and increase productivity~\cite{Kelly2008}.
Automatically generating code is a generic approach where the same generator can be reused to produce many different artifacts according to the varying inputs it receives.
It also provides opportunities to detect errors in the input artifact early on before the generated code is compiled, when the output is source code.

There are many techniques to generate code, such as programmatically, using a meta-object protocol, or aspect-oriented programming.
Since the mid-1990s, \linebreak template-based code generation (TBCG) emerged as an approach requiring less effort for the programmers to develop code generators.
Templates favor reuse following the principle of \emph{write once, produce many}. The concept was heavily used in web designer software (such as Dreamweaver) to generate web pages and Computer Aided Software Engineering (CASE) tools to generate source code from UML diagrams.
Many development environments started to include a template mechanism in their framework such as Microsoft Text Template Transformation Toolkit (T4)\footnote{\url{https://msdn.microsoft.com/en-us/library/bb126445.aspx}} for .NET and Velocity\footnote{\url{http://velocity.apache.org/}} for Apache.

Model-driven engineering (MDE) has advocated the use of model-to-text transformations as a core component of its paradigm~\cite{Kleppe2003}.
TBCG is a popular technique in MDE given that they both emphasize abstraction and automation.
MDE tools, such as Acceleo\footnote{\url{http://www.eclipse.org/acceleo/}} and Xpand\footnote{\url{http://wiki.eclipse.org/Xpand}}, allow developers to generate code from high-level models without worrying on how to parse and traverse input models.
We can find today TBCG applied in a plethora of computer science and engineering research.


The software engineering research community has focused essentially on primary studies proposing new TBCG techniques, tools and applications. However, to the best of our knowledge, there is no classification, characterization, or assessment of these studies available yet.
Therefore, in this paper, we conducted a systematic mapping study of the literature in order to understand the trends, identify the characteristics of TBCG, assess the popularity of existing tools, and determine the influence that MDE has had on TBCG. We are interested in various facets of TBCG, such as characterizing of the templates, of inputs and outputs, along with the evolution of the amount of publications using TBCG over the past 15 years.

The remainder of this paper is organized as follows. In \Sect{sec:relatedwork}, we introduce the necessary background on TBCG and discuss related work. In \Sect{sec:methods}, we elaborate on the methodology we followed for this systematic mapping study. We present the results of the paper selection phase in \Sect{sec:selection}. The following sections report the results. First in \Sect{sec:trends}, we describe the trends and evolution of TBCG over the past 15 years. Then in \Sect{sec:characteristics}, we discuss the characteristics of TBCG according to our classification scheme. We explain the relationships between the different facets in \Sect{sec:correlations}. Next in \Sect{sec:tools}, we discuss how TBCG tools have been used in primary studies. In \Sect{sec:mde}, we discuss about the relation between MDE and TBCG. In \Sect{sec:discussion}, we answer our research questions and discuss limitations of the study. Finally, we conclude in \Sect{sec:conclusion}.

%% file: relatedwork.tex
\section{Background and Related Work}\label{sec:relatedwork}

In this section, we review the notion of code generation and introduce TBCG.
We also briefly outline MDE principles to better understand its relationship with TBCG.
Finally we discuss related work on systematic mapping studies in general and secondary studies about code generation.

\subsection{Code Generation}

In this paper, we view code generation as in automatic programming~\cite{Rich1988} rather than compilers.
The underlying principle of automatic programming is that a user defines what he expects from the program and the program should be automatically generated by a software without any assistance by the user.
This generative approach is different from a compiler approach.

As Blazer~\cite{Blazer1985} states, there are many advantages to code generation.
The effort of the user is reduced as he has fewer lines to write: specifications are shorter than the program that implements them.
Specifications are easier to write and to understand for a user, given that they are closer to the application and domain concepts.
Writing specifications is less error-prone than writing the program directly, since the expert is the one who writes the specification rather than another programmer.

These advantages are in fact the pillar principles of MDE and domain-specific modeling.
Floch\etal{}~\cite{Floch2011} observed many similarities between MDE and compilers research and principles.
Thus, it is not surprising to see that many, though not exclusively, code generation tools came out of the MDE community.
The advantages of code generation should be contrasted with some of its limitations.
For example, there are issues related to integration of generated code with manually written code and to evolving specifications that require to re-generate the code~\cite{Stahl2006}.
Sometimes, relying too much on code generators may produce an overly general solution that may not necessarily be optimal for a specific problem.

\subsection{Code Generation in the Context of MDE}\label{sec:codeg}

MDE is a software development approach that uses abstraction to bridge the gap between the problem space and the software implementation~\cite{Stahl2006}.
To bridge the gap between the application domain and the solution domain, MDE uses models to describe complex systems at multiple levels of abstraction, as well as automated support for transforming and analyzing models. This separation allows the description of key intellectual assets in a way that is not coupled to specific programming languages or target platforms.

Domain-specific modeling (DSM)~\cite{Gray2007} is a branch of MDE that allows models to be manipulated at the level of abstraction of the application domain the model is intended for, rather than at the level of computing. In DSM, domain experts can create models that describe some computational need using abstractions and notations that match their own domain of expertise. Thus, end-users who do not possess the skills needed to write computer programs using traditional languages (like Java or C++) can describe their solution in a more familiar language.

In MDE parlance, models represent abstractions of a real system, capturing some of its essential properties.
A model conforms to a metamodel, which defines the abstract syntax and static semantics of the modeling language.
This language can either be a domain-specific language (DSL) or a general purpose language like UML.
Developer manipulates models by means of model transformation.
Transformations can have different purposes~\cite{Lucio2014}, such as translating, simulating, or refining models.
One particular kind of model transformation is devoted to code generation with model-to-text transformations~\cite{Czarnecki2006}.

A common workflow in MDE is to produce a program without the need of programming~\cite{Kleppe2003}.
Modelers first describe the high-level level system in a computation-indepen\-dent model.
This is then evolved into a domain-specific platform-independent model.
This model is in turn refined with platform-specific concepts from the target framework of the final application.
The platform-specific model is then synthesized to the source code of the program using a dedicated model-to-text transformation tool~\cite{Rose2012}.
Model-to-text transformations are used to implement code, generate documentation, serialize models, or visualize and explore models.
We refer to~\cite{Jorges2013} for a history of code generation and an in-depth explanation of its role in MDE.

\subsection{Code Generation Techniques}\label{sec:cg-techniques}

As briefly outlined in~\cite{Jorges2013} and in~\cite{Czarnecki2006}, there are many techniques that can be used to generate code.
We briefly outline the main ones here.

\begin{description}
	\item[Visitor based] approaches consist of programmatically traversing the internal representation of the input, while relying on an API dedicated to manipulate the input and to write the output to a text stream. This is used in~\cite{Bonta2009}.
	\item[Meta-programming] is a language extension approach, such as reflection or using a meta-object protocol. For example, in OpenJava~\cite{Tatsubori2000}, a Java meta-program creates a Java file, compiles it on the fly, and loads the generated program in its own run-time.
	\item[Aspect-oriented programming] is a language composition approach that weaves code fragments at specific locations in a program to add new concerns to the existing code. This is used in~\cite{Lohmann2004}.
	\item[In-line generation] relies on a preprocessor that generates additional code to the existing one, such as with the C++ standard template library or C macro preprocessor instructions. An example is available in~\cite{Beckmann2004}.
	\item[Code annotations] are added in-line to existing code and is internally transformed into more expanded code. Examples include JavaDoc and attributes in C\#. This approach is used in \cite{Cordoba2015}.
	\item[Template based] is described below.
\end{description}

\subsection{Template-based Code Generation} \label{sec:tbcg}

The literature agrees on a general definition of model-to-text code generation~\cite{Czarnecki2006} and on templates.
J{\"o}rges~\cite{Jorges2013} identifies three components in TBCG: the data, the template, and the output.
However, there is another component that is not mentioned which is the meta-information the generation logic of the template relies on.
Therefore, we conducted this study according to the following notion of TBCG.

\begin{figure}[h]
	\centering
	\includegraphics[width=\linewidth]{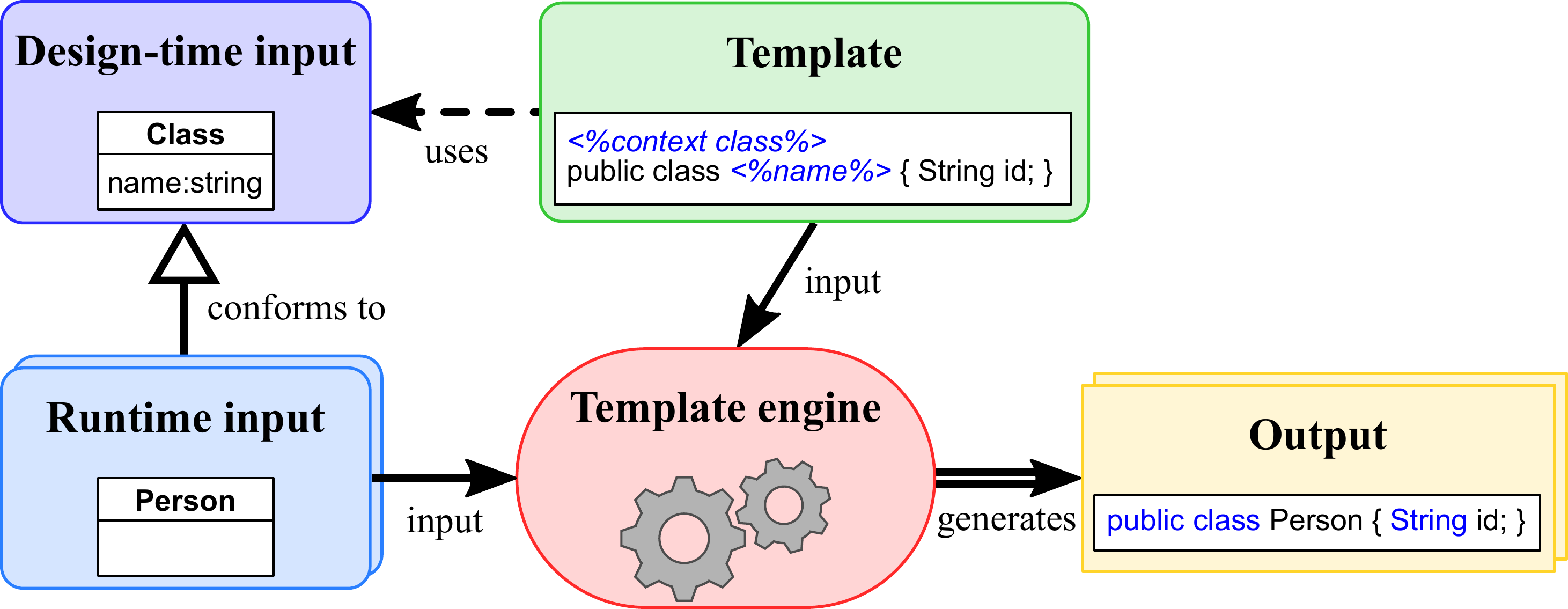}
	\caption{Components of TBCG}
	\label{fig:components}
\end{figure}
\Fig{fig:components} summarizes the main concepts of TBCG.
We consider TBCG as a synthesis technique that uses templates in order to produce a textual artifact, such as source code, called the \emph{output}.
A template is an abstract and generalized representation of the textual output it describes.
It has a \emph{static part}, text fragments that appear in the output ``as is''.
It also has a \emph{dynamic part} embedded with splices of meta-code that encode the generation logic.
Templates are executed by the \emph{template engine} to compute the dynamic part and replace meta-codes by static text according to \emph{run-time input}.
The \emph{design-time input} defines the meta-information which the run-time input conforms to.
The dynamic part of a template relies on the design-time input to query the run-time input by filtering the information retrieved and performing iterative expansions on it.
Therefore, TBCG relies on a design-time input that is used to define the template and a run-time input on which the template is applied to produce the output.
For example, a TBCG engine that takes as run-time input an XML document relies on an XML schema as design-time input.
Definition~\ref{def:tbcg} summarizes our definition of TBCG.
\begin{defn}\label{def:tbcg}
	A synthesis technique is a TBCG if and only if it specifies a set of templates, assumes a design-time input, requires run-time inputs, and produces textual output.
\end{defn}
For example, the work in \cite{Jugel2011} generates a C\# API from Ecore models using Xpand. 
According to Definition~\ref{def:tbcg}, the templates of this TBCG example are Xpand templates, the design-time input is the metamodel of Ecore, the run-time input is an Ecore model, and the output is a C\# project file and C\# classes.
%

\subsection{Literature Reviews on Code Generation} \label{sec:sms}

In evidence-based software engineering~\cite{Kitchenham2004}, a systematic literature review is a secondary study that reviews primary studies with the aim of synthesizing evidence related to a specific research question.
Several forms of systematic reviews exist depending on the depth of reviewing primary studies and on the specificities of research questions.
Unlike conventional systematic literature reviews that attempt to answer a specific question, a systematic mapping studies (SMS) aim at classifying and performing a thematic analysis on a topic~\cite{Kitchenham2011}.
SMS is a secondary study method that has been adapted from other disciplines to software engineering in~\cite{Brereton2007} and later evolved by Petersen\etal{} in~\cite{Petersen2008}.
A SMS is designed to provide a wide overview of a research area, establish if research evidence exists on a specific topic, and provide an indication of the quantity of the evidence specific to the domain.

Over the years, there have been many primary studies on code generation. However, we could not find any secondary study on TBCG explicitly. Still, the following are closely related secondary studies.

Mehmood\etal{}~\cite{Mehmood2013} performed a SMS regarding the use of aspect-oriented modeling for code generation, which is not based on templates. They analyzed 65 papers mainly based on three main categories: the focus area, the type of research, and the type of contribution. The authors concluded that this synthesis technique is still immature. The study shows that no work has been reported to use or evaluate any of the techniques proposed.

Gurunule\etal{} \cite{Gurunule2015} presented a comparison of aspect orientation and MDE techniques to investigate how they can each be used for code generation. The authors found that further research in these areas can lead to significant advancements in the development of software systems. Unlike Mehmood\etal{}~\cite{Mehmood2013}, they did not follow a systematic and repeatable process.

Dominguez\etal{}~\cite{Domiguez2012} performed a systematic literature review of studies that focus on code generation from state machine specifications. The study is based on a set of 53 papers, which have been classified into two groups: pattern-based and not pattern-based.
The authors do not take template-based approaches into consideration.

Batot\etal{}~\cite{Batot2016} performed a SMS on model transformations solving a concrete problem that have been published in the literature. They analyzed 82 papers based on a classification scheme that is general to any model transformation approach, which includes model-to-text transformations. They conclude that concrete model transformations have been pulling out from the research literature since 2009 and are being considered as development tasks. They also found that 22\% of their corpus solve concrete problems using refinement and code synthesis techniques. Finally, they found that research in model transformations is heading for a more stable and grounded validation.

There are other studies that attempted to classify code generation techniques. However, they did not follow a systematic and repeatable process.
For example, Czarnecki\etal{}~\cite{Czarnecki2006} proposed a feature model providing a terminology to characterize model transformation approaches. They distinguished two categories for model-to-text approaches: those that are visitor-based and those that are template-based; the latter being in line with Definition~\ref{def:tbcg}. The authors found that many new approaches to model-to-model transformation have been proposed recently, but relatively little experience is available to assess their effectiveness in practical applications. 

Rose\etal{}~\cite{Rose2012} extended the feature model of Czarnecki \etal{} to focus on template-based model-to-text transformation tools. Their classification is centered exclusively on tool-dependent features. Their goal is to help developers when they are faced to choose between different tools.
This study is close to the work of Czarnecki in \cite{Czarnecki2006} but focuses only on a feature model for M2T.
The difference with our study is that it focuses on a feature diagram and deals with tool-dependent features only.

%% file: methods.tex
\section{Research Methods} \label{sec:methods}

\begin{figure*}
	\centering
	\includegraphics[width=\linewidth]{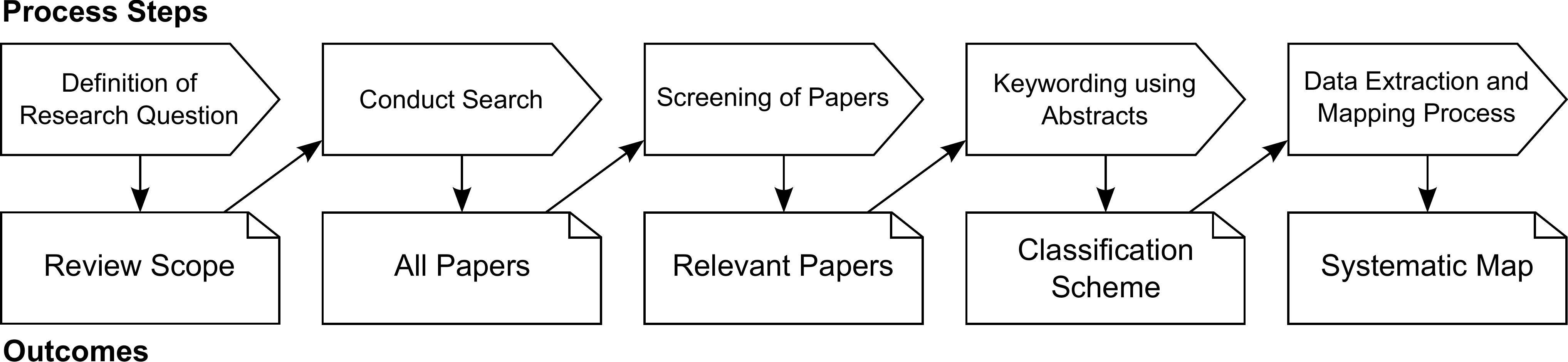}
	\caption{The systematic mapping process we followed}
	\label{fig:process}
\end{figure*}
In order to analyze the topic of TBCG, we conducted a SMS following the process defined by Petersen\etal{} in~\cite{Petersen2008} and summarized in \Fig{fig:process}.

\subsection{Objectives}

The objective of this study is to obtain an overview of the current research in the area of TBCG and to characterize the different approaches that have been developed. We defined four research questions to set the scope of this study:
\begin{enumerate}
	\item \textbf{What are the trends in template-based code generation?} We are interested to know how this technique has evolved over the years.
	\item \textbf{What are the characteristics of template-based code generation approaches?} We want to identify major characteristics of this techniques and their tendencies.
	\item \textbf{To what extent are template-based code generation tools being used?} We are interested in identifying popular tools and their uses.
	\item \textbf{What is the place of MDE in template-based code generation?} We seek to determine whether and how MDE has influenced TBCG.
\end{enumerate}

\subsection{Selection of Source}\label{sec:query}

We delimited the scope of the search to be regular publications that mention TBCG as at least one of the approaches used for code generation and published between 2000--2015. Therefore, this includes publications where code generation is not the main contribution. For example, Buchmann\etal{}~\cite{Buchmann2013} used TBCG to obtain ATL code while their main focus was implementing a higher-order transformation. Given that not all publications have the term ``code generation'' in their title, we formulated a query that retrieves publications based on their title, abstract, or full text (when available) mentioning ``template'' and ``code generation'', their variations, and synonyms.
%
The used query was validated with a sample of 100 pre-selected papers we knew should be included.

\subsection{Screening Procedure} \label{sec:screening}

Screening is the most crucial phase in a SMS~\cite{Petersen2008}.
We followed a two-stage screening procedure: automatic filtering, then title and abstract screening.
In order to avoid the exclusion of papers that should be part of the final corpus, we followed a strict screening procedure.
With four reviewers at our disposal, each article is screened by at least two reviewers independently.
When both reviewers of a paper disagree upon the inclusion or exclusion of the paper, a physical discussion is required.
If the conflict is still unresolved, an additional senior reviewer is involved in the discussion until a consensus is reached.
To determine a fair exclusion process, a senior reviewer reviews a sample of no less than 20\% of the excluded papers at the end of the screening phase, to make sure that no potential paper is missed.

\subsubsection{Inclusion criteria}

A paper is included if it explicitly indicates the use of TBCG or if it proposes a TBCG technique. We also include papers if the name of a TBCG tool appears in the title, abstract, or content.

\subsubsection{Exclusion criteria}\label{sec:exclusion}

Results from the search were first filtered automatically to discard records that were outside the scope of this study:
papers not in computer science, not in the software engineering domain, with less than two pages of length (\eg proceedings preface), not peer-reviewed (\eg white papers), not written in English, or not published between the years 2000 and 2015.
Then, papers were excluded through manual inspection based on the following criteria:
\begin{itemize}
	\item \textbf {No code generation.} There is no code generation technique used.
	\item \textbf{Not template-based code generation.} Code generation is mentioned, but the considered technique is not template-based according to Definition~\ref{def:tbcg}.
	\item \textbf{Not a paper.} This exclusion criterion spans papers that were not caught by the automatic filtering. For example, some papers had only the abstract written in English and the content of the paper in another language. Additionally, there were 24 papers where the full text was not accessible online.
\end{itemize}

For the first two criteria, when the abstract did not give enough details about the code generation approach, a quick look at the full text helped clear any doubts on whether to exclude the paper or not. Reviewers were conservative on that matter.

\subsection{Classification Scheme} \label{sec:classification}

There are generally two ways to construct the classification scheme~\cite{Petersen2008,Seriai2014}.
One approach consists of extracting the classification scheme by analyzing the included papers and determining the
important classification properties form the abstract, keywords or content.
Alternatively, one can construct a scheme using the general knowledge of the field.
In our study, we used a hybrid approach in which we combined our general knowledge with the information extracted from the abstracts during the screening phase.
The classification scheme is used to classify all retained papers along different facets that are of interest in order to answer our research questions.
It helps analyzing the overall results and gives an overview of the trends and characteristics of TBCG.
The categories we classified the corpus with are the following:
\begin{description}
\item \textbf{Template style}: We characterize the style of the templates used in code generation approach.
\begin{itemize}
	\item \textbf{Predefined}: This template style is reserved for approaches where the template used for code generation is defined internally to the tool. However, a subset of the static part of the template is customizable to vary slightly the generated output. This is, for example, the case for common CASE tools where there is a predefined template to synthesize a class diagram into a number of programming languages. Nevertheless, the user can specify what language construct to use for association ends with a many cardinality, such as \Code{Array} or \Code{ArrayList} for Java templates.
	\item \textbf{Output-based}: This style covers templates that are syntactically based on the actual target output. In contrast with the previous style, output-based templates offer full control on how the code is generated, both on the static and dynamic parts. The generation logic is typically encoded in meta-code as in the example of \Fig{fig:components}.
	\item \textbf{Rule-based}: In this style, templates focus on computing the dynamic part with the static part being implicit. The template lists declarative production rules that are applied on-demand by the template engine to obtain the final target output. For example, this is used to render the concrete textual syntax from the abstract syntax of a model using a grammar.
\end{itemize}

\item \textbf{Design-time input type}: We characterize the language of the design-time input that is necessary to develop templates.
\begin{itemize}
	\item \textbf{General purpose}: for generic languages reusable across different domains that are not programming languages, such as UML.
	\item \textbf{Domain specific}: for languages targeted for a particular domain, such as the metamodel of a DSL.
	\item \textbf{Schema}: for structured data definitions, such as XML schema definition or database schema.
	\item \textbf{Programming language}: for well-defined programming languages.
\end{itemize}

\item \textbf{Run-time input type}: We characterize the input given to the generator during the execution of a TBCG. Generally, the run-time input is an instance that conforms to the design-time input.
\begin{itemize}
	\item \textbf{General purpose}: for instances of a generic languages, such as the Ecore model of a particular class diagram.
	\item \textbf{Domain specific}: for instances of a domain-specific language, such as a Simulink model.
	\item \textbf{Structured data}: for data that follows a well-defined structure, such as XML.
	\item \textbf{Source code}: when the input is source code implemented in a given programming language.
\end{itemize}

\item \textbf{Output type}: We characterize the artifacts output by the code generator. A paper may be classified in more than one of the following categories.
\begin{itemize}
	\item \textbf{Source code}: for executable code conforming to a specific programming language.
	\item \textbf{Structured data}: for code that is not executable, such as HTML.
	\item \textbf{Natural language}: when plain text is generated.
\end{itemize}

\item \textbf{Tool}: We capture the tool or language used for TBCG. If a tool is not clearly identified in a paper or the TBCG is programmed directly, we classify the tool as \textbf{unspecified}. We consider a tool to be \textbf{popular} when it is used in at 1\% of the papers. Otherwise, we classify it in the \textbf{other} category.

\item \textbf{MDE}: We determine whether the part of the solution where TBCG is applied in the paper follows MDE techniques and principles. A good indication is if the design-time input is a metamodel.

\item \textbf{Context}: We determine where TBCG falls in the overall transformation process of the approach. We already presented a typical workflow in \Sect{sec:codeg}. Code generation is never the first step unless it is \textbf{standalone}. Otherwise, it is either used as an \textbf{intermediate} step or it is the \textbf{last} step of a transformation process.

\item \textbf{Validation}: We categorize how the TBCG approach or the generated output is validated. The validation can be performed against a \textbf{benchmark}, based on a \textbf{user study}, applied on a \textbf{case study}, defined \textbf{formally}, or there is \textbf{no validation}.

\item \textbf{Application scale}: We characterize the scale of the artifact on which the TBCG approach is applied. We distinguish between \textbf{large scale} applications, \textbf{small scale}, or \textbf{no application} when the code generation was not applied on any example.

\item \textbf{Application domain}: We classify the general domain TBCG has been applied on. For example, this includes \textbf{Software engineering}, \textbf{Embedded systems}, \textbf{Compilers}, \textbf{Bio-medicine}, etc.

\item \textbf{Orientation}: We distinguish \textbf{industrial} papers, where at least one author is affiliated to industry, from \textbf{academic} papers otherwise.

\item \textbf{Publication type}: We distinguish papers published in \textbf{conference} proceedings, as \textbf{journal} articles, or \textbf{other} formats such as workshop proceedings or book collections.

\item \textbf{Venue type}: We classify papers based on the where they have been published. We distinguish between general \textbf{software engineering} venues, venues specific to \textbf{MDE}, and all \textbf{other} venue types.
\end{description} 

%% file: selection.tex
\section{Paper Selection} \label{sec:selection}

\Table{tab:selection} summarizes the flow of information through the selection process of this study.
This section explains how we obtained the final corpus of papers to classify and analyze.
\begin{table}[h]
    \centering
	{\small
		\begin{tabular}{lr}
			\toprule
			\textbf{Phase} & \textbf{Number of papers} 	\\
			\midrule
			\textbf{Collection}	\\
			Engineering Village & $4\;043$	\\
			Scopus & 916 \\
			SpringerLink & $2\;368$	\\
			\textit{Initial corpus}	& $\mathit{5\;081}$	\\
			\midrule
			\textbf{Screening}	\\
			Excluded during screening & $4\;544$	\\
			\textit{Included} & \textit{537}	\\
			\midrule
			\textbf{Classification} \\
			Excluded during classification & 97	\\
			\textit{Final corpus} & $\mathit{440}$	\\
			\bottomrule
		\end{tabular}}
		\caption{Evolution of paper corpus during the study process}
		\label{tab:selection}
	\end{table}

\subsection{Paper Collection}

The paper collection step was done in two phases: querying and automatic duplicates removal.
There are several online databases that index software engineering literature.
For this study, we considered three main databases to maximize coverage: \textsc{Engineering Village}\footnote{\url{https://www.engineeringvillage.com/}}, \textsc{Scopus}\footnote{\url{https://www.scopus.com/}}, and \textsc{SpringerLink}\footnote{\url{http://link.springer.com/}}.
The first two cover typical software engineering editors (\textsc{IEEE Xplore}, \textsc{ACM Digital Library}, \textsc{Elsevier}). However, from past experiences \cite{Batot2016}, they do not include all of \textsc{Springer} publications.
We used the search string from \Sect{sec:query} to retrieve all papers from these three databases.
We obtained $7\;527$ candidate papers that satisfy the query and the options of the search stated in \Sect{sec:exclusion}.
We then removed automatically all duplicates using EndNote software.
This resulted in $5\;081$ candidate papers for the screening phase.

\subsection{Screening}

Based on the exclusion criteria stated in \Sect{sec:exclusion}, each candidate paper was screened by at least two reviewers to decide on its inclusion.
To make the screening phase more efficient, we used a home-made tool~\cite{Seriai2014}.
After all the reviewers completed screening the papers they were assigned, the tool calculates an inter-rater agreement coefficient.
In our case, the Cohen’s Kappa coefficient was $0.813$.
This high value shows that the reviewers were in almost perfect agreement.

Among the initial corpus of candidate papers, $4\;547$ were excluded, $510$ were included and $24$ received conflicting ratings.
During the screening, the senior reviewer systematically verifies each set of 100 rejected papers for sanity check. A total of $7$ more papers were included back hence the rejected papers were reduced to $4\;540$.
Almost all cases of conflicts were about a disagreement on whether the code generation technique of a paper was using templates or not.
These conflicts were resolved in physical meetings and 20 of them were finally included for a total of 537 papers and $4\;544$ excluded.

Among the excluded papers, 52\% were rejected because \textit{no code generation} was used.
We were expecting such a high rate because terms such as ``templates'' are used in many other fields, like biometrics.
Also, many of these papers were referring to the C++ standard template library \cite{Liu2006}, which is not about code generation.
We counted 34\% papers excluded because they were \textit{not using templates}.
Examples of such papers are cited in \Sect{sec:cg-techniques}.
Also, more than a quarter of the papers were in the compilers or embedded system domains, where code generation is programmed imperatively rather than declaratively specified using a template mechanism.
Finally, 5\% of the papers were considered as \textit{not a paper}.
In fact, this criterion was in place to catch papers that escaped the automatic filtering from the databases.

\subsection{Eligibility during Classification}\label{sec:eligibility}

Once the screening phase over, we thoroughly analyzed the full text of the remaining 537 papers to classify them according to our classification scheme.
Doing so allowed us to confirm that the code generation approach was effectively template-based according to Definition~\ref{def:tbcg}.
We encountered papers that used multiple TBCG tools: they either compared tools or adopted different tools for different tasks.
We classified each of these papers as a single publication, but incremented the occurrence corresponding to the tools referred to in the paper.
This is the case of~\cite{Fang2006} where the authors use Velocity and XSLT for code generation.
Velocity generates Java and SQL code, while XSLT generates the control code.

We excluded 97 additional papers.
During screening, we detected situations where the abstract suggested the implementation of TBCG, whereas the full text proved otherwise. 
In most of the cases, the meaning of TBCG differed from the description presented in \Sect{sec:tbcg}.
As shown in~\cite{Singh1991} the terms template-based and generation are used in the context of networking and distributed systems.
We also encountered circumstances where the tool mentioned in the abstract requires the explicit use of another component to be considered as TBCG, such as Simulink TLC, as in~\cite{OHalloran2013}.
The final corpus\footnote{
	The complete list of papers is available online at \url{http://www-ens.iro.umontreal.ca/~luhunukl/classification.html}}
considered for this study contains 440 papers.

%% file: trends.tex
\section{Evolution of TBCG} \label{sec:trends}

We start with a thorough analysis of the trends in TBCG in order to answer the first research question.

\subsection{General trend}

\begin{figure}[h]
	\centering
	\includegraphics[width=.7\linewidth]{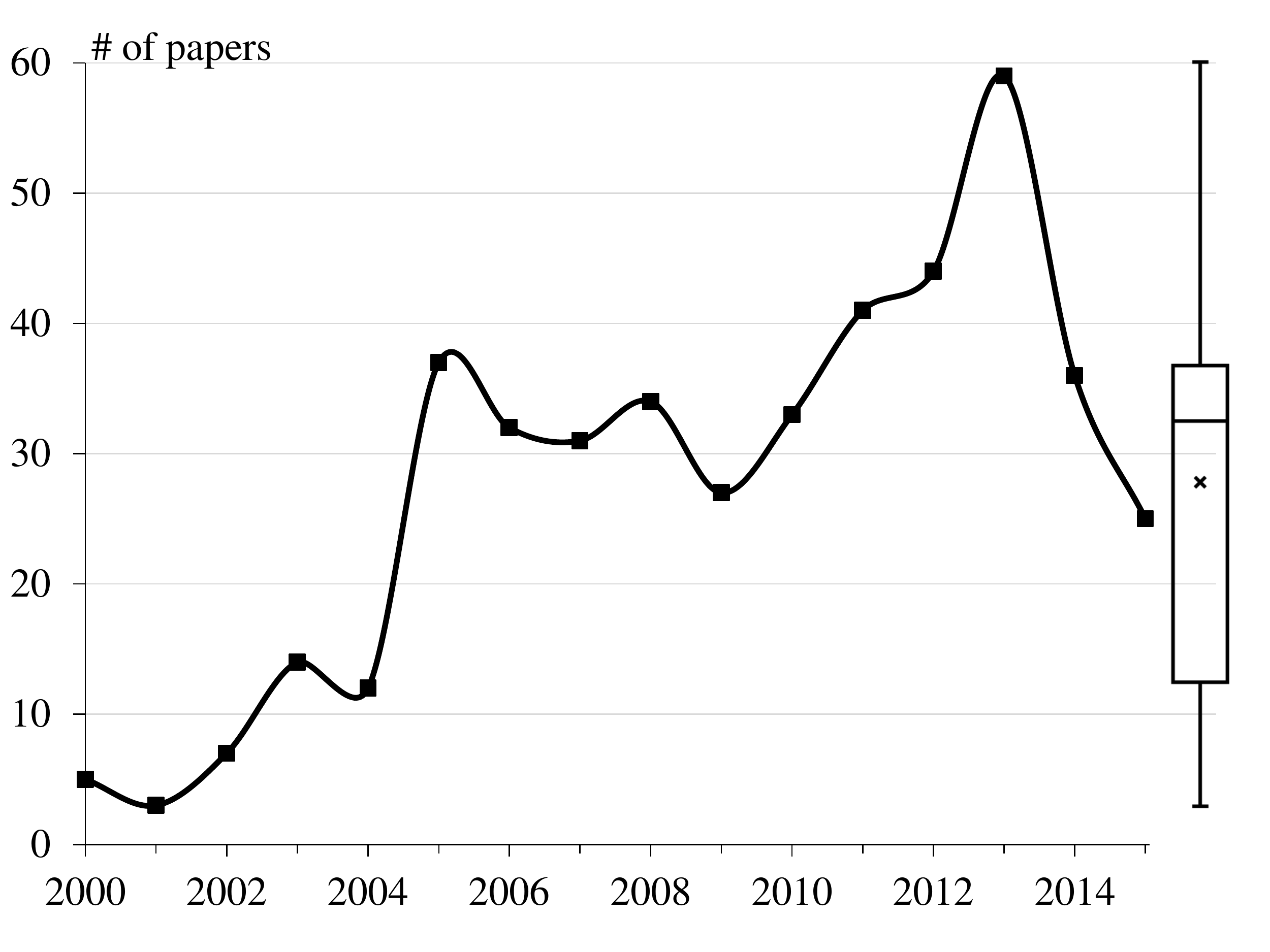}
	\caption{Evolution of papers in the corpus}
	\label{fig:general_trend}
\end{figure}

\begin{table}
\centering
{\tiny
\begin{tabular}{lrlc}
	\toprule
	\textbf{Venue} & \multicolumn{2}{c}{\textbf{Venue \& Publication type}} & \textbf{\# Papers} \\
	\midrule
	Model Driven Engineering Languages and Systems (\textsc{Models}) & MDE & Conference & 26 \\
	Software and Systems Modeling (\textsc{Sosym}) & MDE & Journal & 23 \\
	European Conference on Modeling Foundations and Applications (\textsc{Ecmfa}) & MDE & Conference & 17 \\
	Generative and Transformational Techniques in Software Engineering (\textsc{Gttse}) & Soft. eng. & Conference & 11 \\
	Generative Programming: Concepts \& Experience (\textsc{Gpce}) & MDE & Conference & 8	\\
	International Conference on Computational Science and Applications (\textsc{Iccsa}) & Other & Conference & 8	\\
	Software Language Engineering (\textsc{Sle})  &  MDE & Conference & 7	\\
	International Conference on Web Engineering (\textsc{Icwe})  &  Other & Conference & 6	\\
	Leveraging Applications of Formal Methods, Verification and Validation (\textsc{Isola}) &	Other & Conference & 5	\\
	Distributed Applications and Interoperable Systems (\textsc{Dais})  &	Other & Conference & 5	\\
	Evaluation of Novel Approaches to Software Engineering (\textsc{Enase}) &	Soft. eng. & Conference & 5	\\
	\bottomrule
\end{tabular}}
\caption{Most popular venues}
\label{tab:venues}
\end{table}
\Fig{fig:general_trend} reports the number of papers from the final corpus per year.
The general trend indicates that the number of publications with at least one template-based code generation method started increasing in 2002 to reach a first local maximum in 2005 and then remained relatively constant until 2012.
This increase coincides with the early stages of MDE and the first edition of the MODELS conference, previous called UML conference.
This is a typical trend where a research community gets carried away by the enthusiasm of a new potentially interesting domain, which leads to more publications.
However, this does not represent the most prolific period for TBCG.
In fact, in 2013 we notice a significant peak with twice the average numbers of publications observed in the previous years.
\Fig{fig:general_trend} then shows a decreasing trend in the last two years.

To explain this unusual peak, we resorted to statistical methods.
The analysis shows that the coefficient of variability is very high (57\%) which indicates an outlier situation.
After eliminating outliers using the modified Thompson Tau test, the remaining years are 2005--2009 and 2014.
When we only consider these years, the average is 34 papers per year and a coefficient of variability at 11\%.
Therefore, this sudden isolated peak in 2013 is the result of a special event or popularity of TBCG.

To explain the decrease observed in 2015, we went back to the initial corpus from the databases.
We noticed that we were only able to collect 24\% fewer papers in 2015 compared to the previous five years where the online database query returned 327 publications on average, before title elimination.
All data was collected until February 2016, therefore this fewer number of publications may have occurred because not all 2015 papers had been indexed by the databases we queried at that time yet.
Therefore, the decrease in the amount of papers published after 2013 should not be interpreted as a decline in interest in TBCG, but that some event happened around 2013 which boosted publications, and then it went back to the steady rate of publication as previous years.

\subsection{Publications and venues}

We analyzed the papers based on the type of publication and the venue of their publication.
Only 23\% and 21\% of the papers were published respectively in MDE and software engineering venues, while the majority (56\%) were published in other venues.
\Table{tab:venues} shows the most popular venues that have at least five papers from the final corpus.
These top venues account for just more than a quarter of the total number of publications.
Among them, MDE venues account for 67\% of the papers.
\textsc{Models}\footnote{We grouped the \textsc{Uml} conference with \textsc{Models}.}, \textsc{Sosym}, and \textsc{Ecmfa}\footnote{We grouped the \textsc{Ecmda-fa} conference with \textsc{Ecmfa}.} are the three most popular venues with a total of 66 publications between them.
This is very significant given that the average is only 1.7 paper per venue with a standard deviation of 2.61.
Also, 53\% of venues had only one paper using TBCG, which is the case for most of the other venues.

Papers published at MDE venues increased gradually to reach a peak in 2013, whereas software engineering venues experienced a later start with the first set of papers being published in 2005.
The number of papers published at other venues gradually increased until 2007, then experience a brief drop, before gradually increasing again until 2012.
Therefore, although mainly influenced by MDE venues, the unexpected peak in 2013 is the result of an accumulation of the small variations among all venues.
For example, there were typically 2-3 papers published at \textsc{Models}, but in 2013 there were 4.

As for the publication type, conference publications have been dominating at 65\%.
After 2013, this number dropped by 64\%. following the general evolution trend.
Journal article account for 21\% of all papers.
Interestingly, we notice that the evolution of journal articles remained constant with an average of 10.5 papers per year since 2010.
Therefore, the overall observed decrease after 2013 is due to the statistical influence of conference papers, whereas journals papers using TBCG have retained the same popularity for the past 5 years.
Nevertheless, because of the typical one to two years lag between the submission and the publication of an article in a journal, further data from 2016-2017 would be needed.

%% file: characteristics.tex
\section{Characteristics of Template-Based Code Generation} \label{sec:characteristics}

We examine the characteristics of TBCG using the classification scheme presented in \Sect{sec:classification}.

\subsection{Template style} \label{templatestyle}

\begin{figure}[h]
	\centering
	\includegraphics[width=.7\linewidth,trim={0 0 0 13cm},clip]{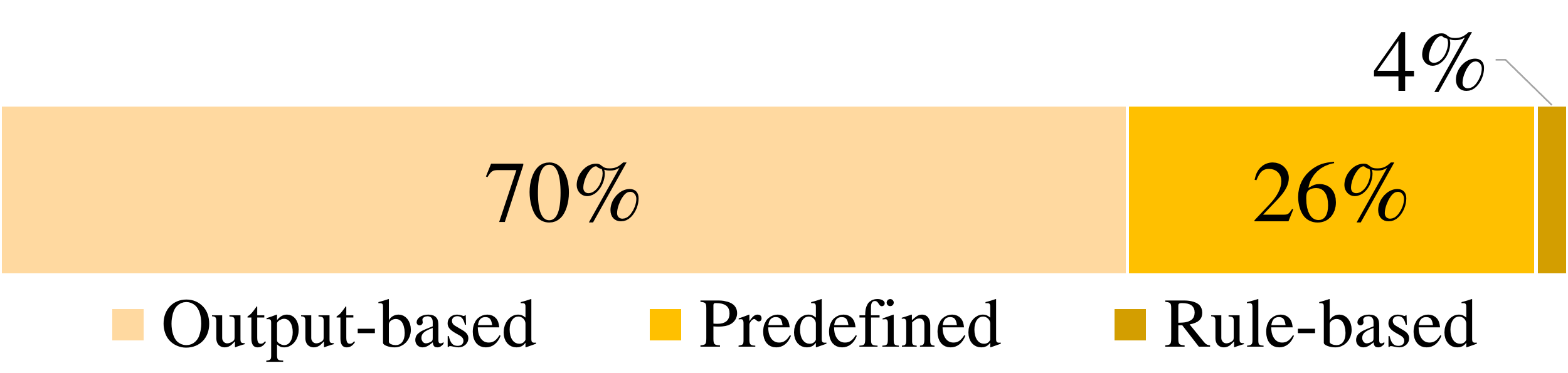}
	\caption{Distribution of template style facet}
	\label{fig:template}
\end{figure}
\begin{figure}
	\centering
	\includegraphics[width=.7\linewidth]{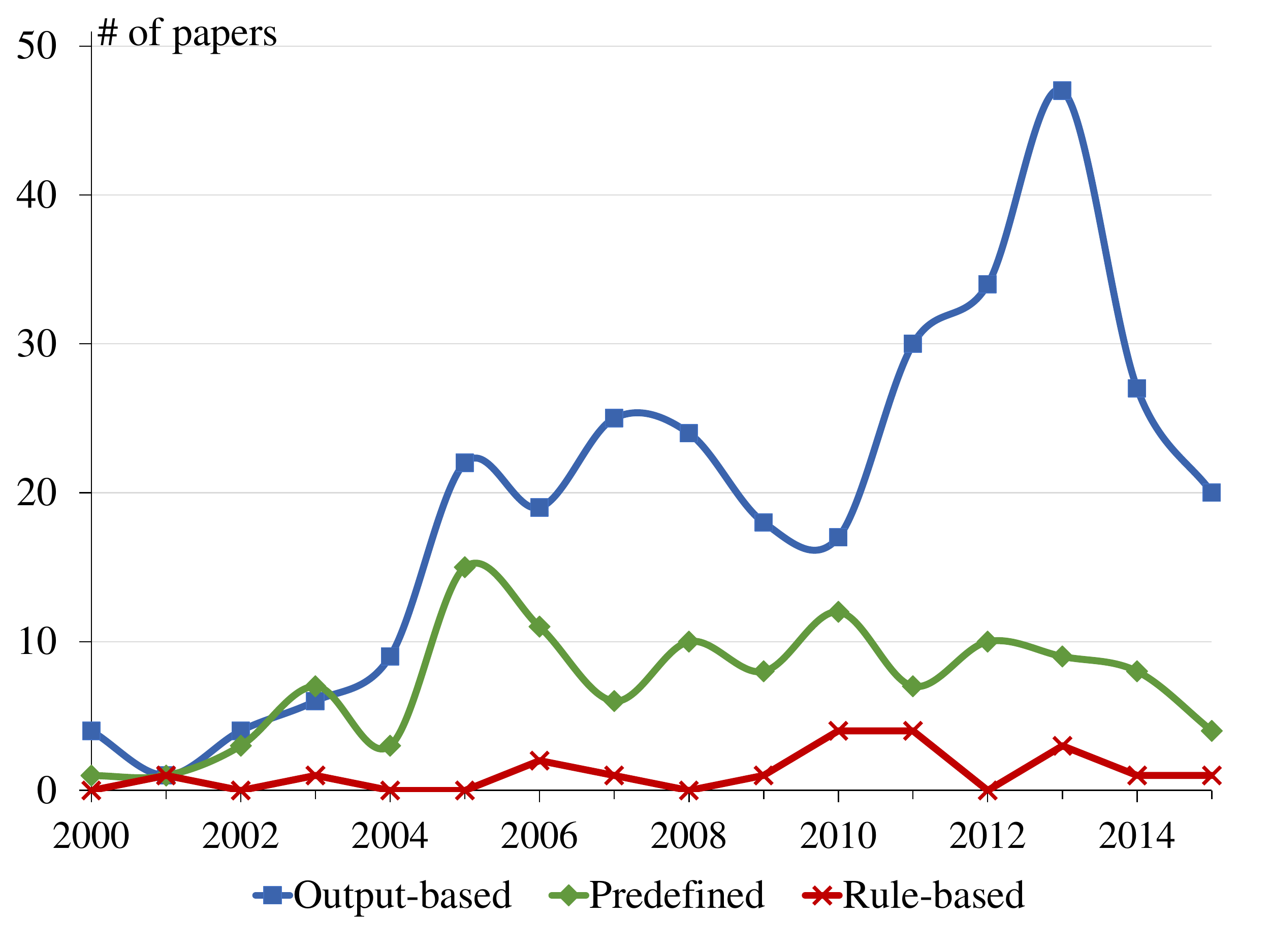}
	\caption{Template style evolution}
	\label{fig:template-trend}
\end{figure}

As the stacked bar chart in \Fig{fig:template} illustrates, the vast majority of the publications follow the \textbf{output-based} style.
This consists of papers like~\cite{Dahman2010}, where Xpand is used to generate workflow code used to automate modeling tools.
There, it is the final output target text that drives the development of the template.
This high score is expected since output-based style is the original template style for TBCG as depicted in \Fig{fig:template-trend}.
This style has always been the most popular style since 2000.

The \textbf{predefined} style is the second most popular.
Most of these papers generate code using a CASE tool, such as~\cite{Gessenharter2008} that uses Rhapsody to generate code to map UML2 semantics to Java code with respect to association ends.
Apart from CASE Tools, we also classified papers like~\cite{Valderas2006} as predefined style since the output code is already fixed as HTML and the programmer uses the tags to change some values based on the model.
There is no other action that can be performed to further customize the final code.
Each year, around 28\% of the papers were using the predefined style, except for a peak of 39\% in 2005, given the popularity of CASE tools then.

We found 19 publications that used \textbf{rule-based} style templates.
This includes papers like~\cite{Hemel2010} which generates Java code with Stratego from a DSL.
A possible explanation of such a low score is that this is the most difficult template style to implement.
It had a maximum of two papers per year throughout the study period.

\subsection{Input type} \label{sec:designtime}

\begin{figure}[h]
	\centering
	\subfloat[Distribution of design-time input type facet]{\includegraphics[width=.7\linewidth,trim={0 0 0 13cm},clip]{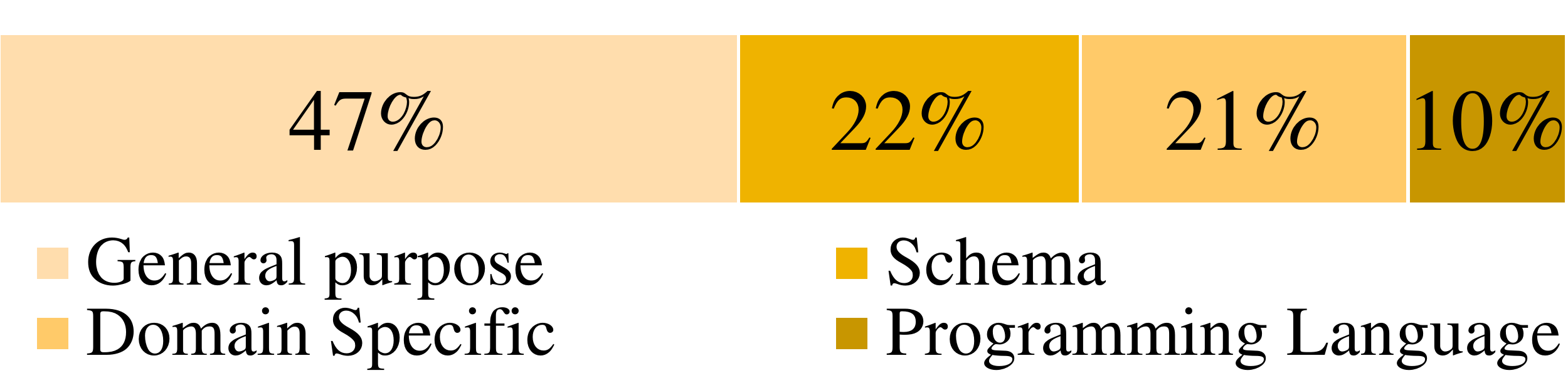} \label{fig:design-time}}\hfill{}
	\subfloat[Distribution of runtime input type facet]{\includegraphics[width=.7\linewidth,trim={0 0 0 13cm},clip]{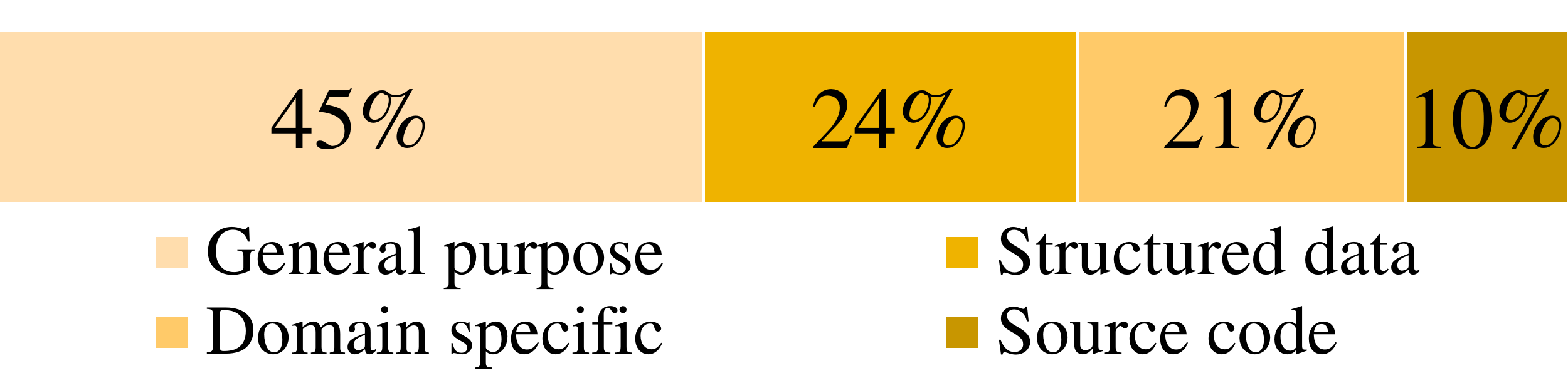}  \label{fig:runtime}}
	\caption{Input types}
\end{figure}
\textbf{General purpose languages} account for almost half of the design-time input of the publications, as depicted in \Fig{fig:design-time}.
UML diagrams, which are used as metamodels for code generation, are the most used for 87\% of these papers.
This is the case in~\cite{Dahman2010} where a class diagram is provided an design-time input to generate workflow.
Other general purpose languages that were used are, for example, the architecture analysis and design language (AADL)~\cite{Brun2008} and feature diagrams \cite{Buckl2005}.
The \textbf{schema} category comes second with 22\% of the papers.
For example, a database schema is used as input at design-time in~\cite{Li2012} to generate Java for a system that demonstrates that template can improve software development.
Also, an XML schema is used in~\cite{Gopinath2011} as design-time input to produce C programs in order to implement an approach that can efficiently support all the configuration options of an application in embedded systems.
\textbf{DSLs} are almost at par with schemata.
They have been gaining popularity and gradually reducing the gap with general purpose languages.
For example in~\cite{Buckl2007}, a custom language is given as the design input in order to generate C and C++ to develop a TBCG approach dedicated to real-time systems.
The least popular design-time input type is \textbf{programming language}.
This includes papers like~\cite{Fischer2014} where T4 is used to generate hardware description (VHDL) code for configurable hardware.
In this case, the input is another program on which the template depends.

\begin{figure}[h]
	\centering
	\includegraphics[width=.7\linewidth]{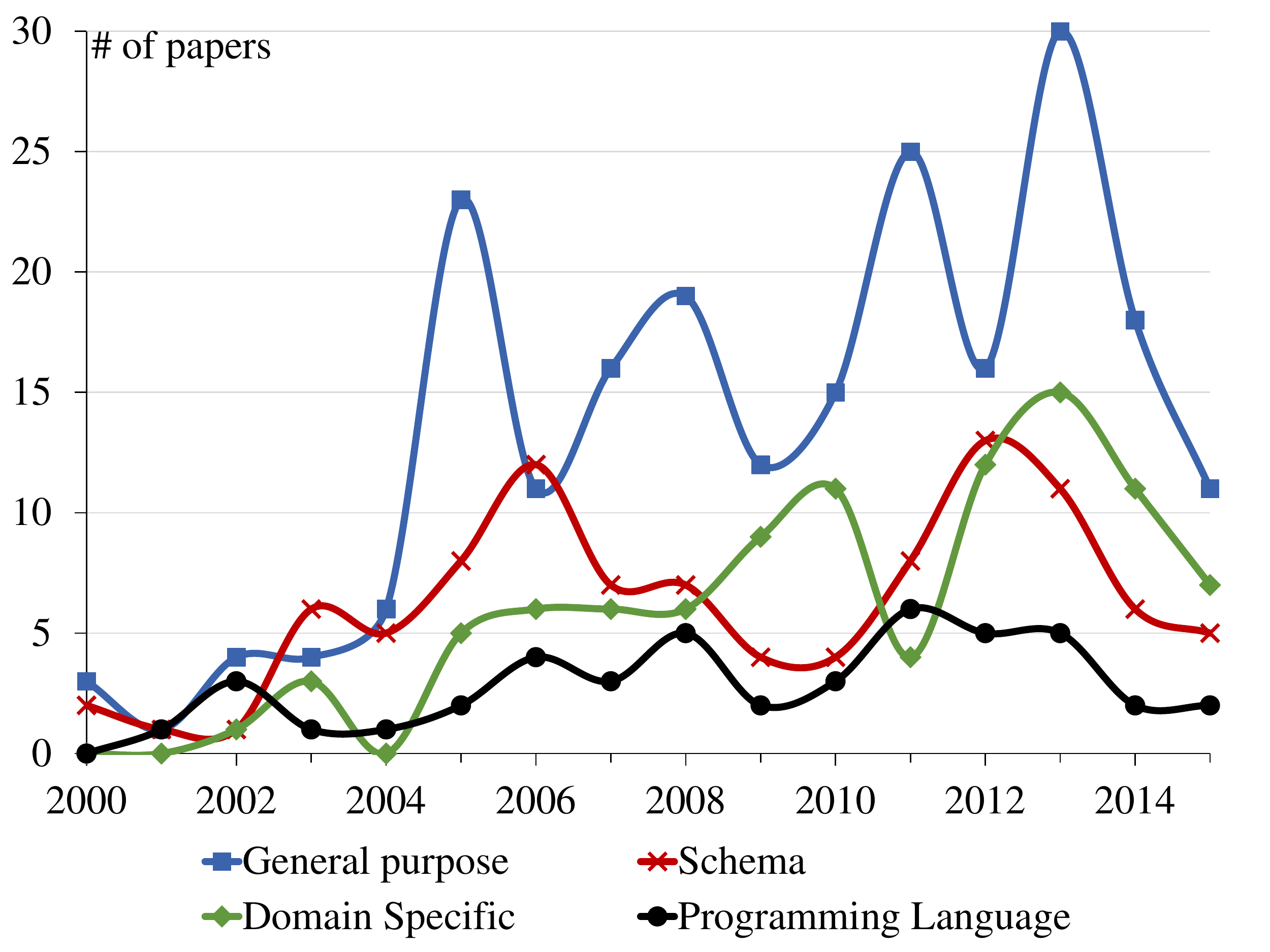}
	\caption{Design-time input evolution}
	\label{fig:designtime-trend}
\end{figure}
Over the years, the general purpose category has dominated the design-time input facet, as depicted in \Fig{fig:designtime-trend}.
2003 and 2006 were the only exceptions where schema obtained slightly more publications.
We also notice a shift from schema to domain-specific design-time input types.
Domain-specific input started increasing in 2009 but never reached the same level as general purpose.
Programming language input maintained a constant level,with an average of 1\% per year.
Interestingly, in 2011, there were more programming language used than DSLs.

Run-time input shown in \Fig{fig:runtime} follows the same trend as design-time input.
This is expected since run-time input is an instance of design-time input.

\subsection{Output type}

\begin{figure}[h]
	\centering
	\includegraphics[width=.7\linewidth,trim={0 0 0 13.2cm},clip]{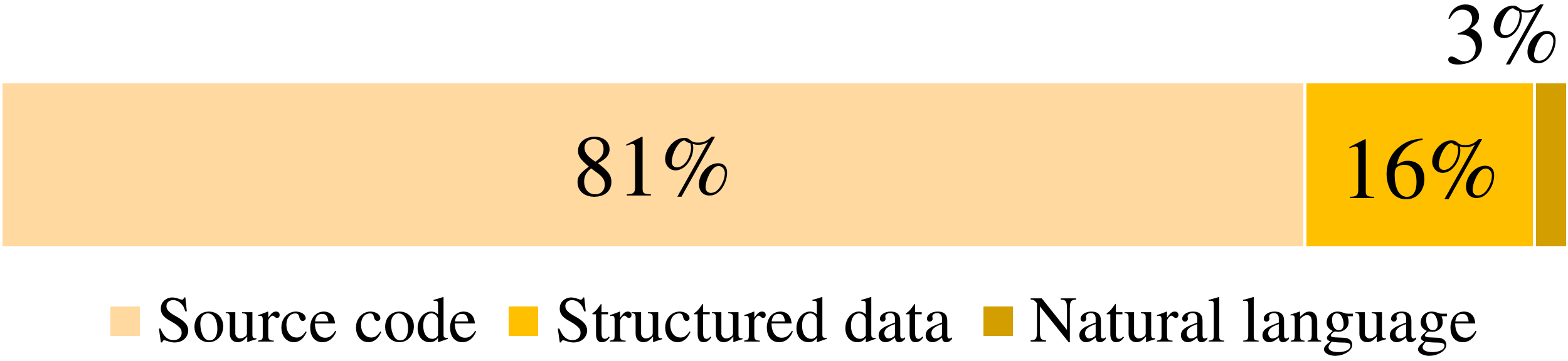}
	\caption{Distribution of output type facet}
	\label{fig:output}
\end{figure}
\Fig{fig:output} shows the distribution of output type facet.
An overwhelming majority of the papers use TBCG to generate \textbf{source code}.
This includes papers like \cite{Chen2010} where Java code is generated an adaptable access control tool for electronic medical records.
Java and C are the most targeted programming languages with respectively 67\% and 20\% of the time.
Writing a program manually often requires proved abilities especially with system and hardware languages, such as VHDL~\cite{Brox2013}.
This is why 10\% of these papers generate low level source codes.
Generation of \textbf{structured data} includes TBCG of mainly XML and HTML files.
For example \cite{Fraternali2009} produces both HTML and XML as parts of the web component to ease regression testing.
Interestingly, we were able to find 12 papers that generate \textbf{natural language} text (in English).
This is surprising given the query string we used.
For example in~\cite{Sridhara2011}, the authors present an automatic technique for identifying code fragments that implement high level abstractions of actions and expressing them as a natural language description.
In addition, we found that around 5\% of the papers generate combinations of at least two output types.
This includes papers such as~\cite{Vokavc2005} that generate both C\# and HTML from a domain specific model and~\cite{Dahman2010} that produce Java as well as natural language text for a system that provides workflow and automation tools for modeling.

Structured data and natural language output remained constant over the years, unlike source code which follows the general trend.

\subsection{Application scale}

\begin{figure}[h]
	\centering
	\includegraphics[width=.7\linewidth,trim={0 0 0 14cm},clip]{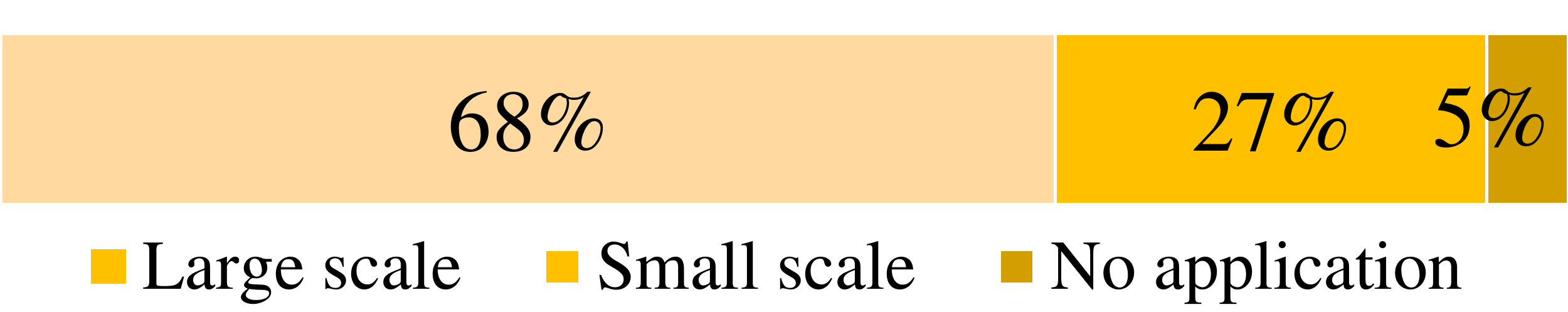}
	\caption{Distribution of application scale facet.}
	\label{fig:applicationscale}
\end{figure}
As depicted in \Fig{fig:applicationscale}, most papers applied TBCG on \textbf{large scale} examples.
This result indicates  that TBCG is a technique that scales with larger amounts of data.
This includes papers like \cite{Roychoudhury2011} that uses Acceleo to generate hundreds of lines of source code to implement an approach that uses models to capture the concepts of various aspect-oriented programming language constructs at a metamodeling level. 
\textbf{small scale}  obtains 26\% of the papers.
This is commonly found in research papers that only need a small and simple example to illustrate their solution.
This is the case in \cite{Hoisl2013} in which a small concocted example shows the generation process with the Epsilon Generation Language (EGL)\footnote{\url{http://www.eclipse.org/epsilon/doc/egl/}}.
\textbf{No application} was used in 5\% of the publications.
This includes papers like \cite{Ecker2014} where authors just mention that code synthesis is performed using a tool named Mako-template.
Even though the number of publications without an actual application is very low, this demonstrates that some authors have still not adopted good practice to show an example of the implementation.
This is important especially when the TBCG approach is performed with a newly developed tool.
While large scale applications follow the general trend of papers, the other two categories remained constant over the years.

\subsection{Validation}

\begin{figure}[h]
	\centering
	\includegraphics[width=0.5\linewidth,trim={0 0 3.6cm 0},clip]{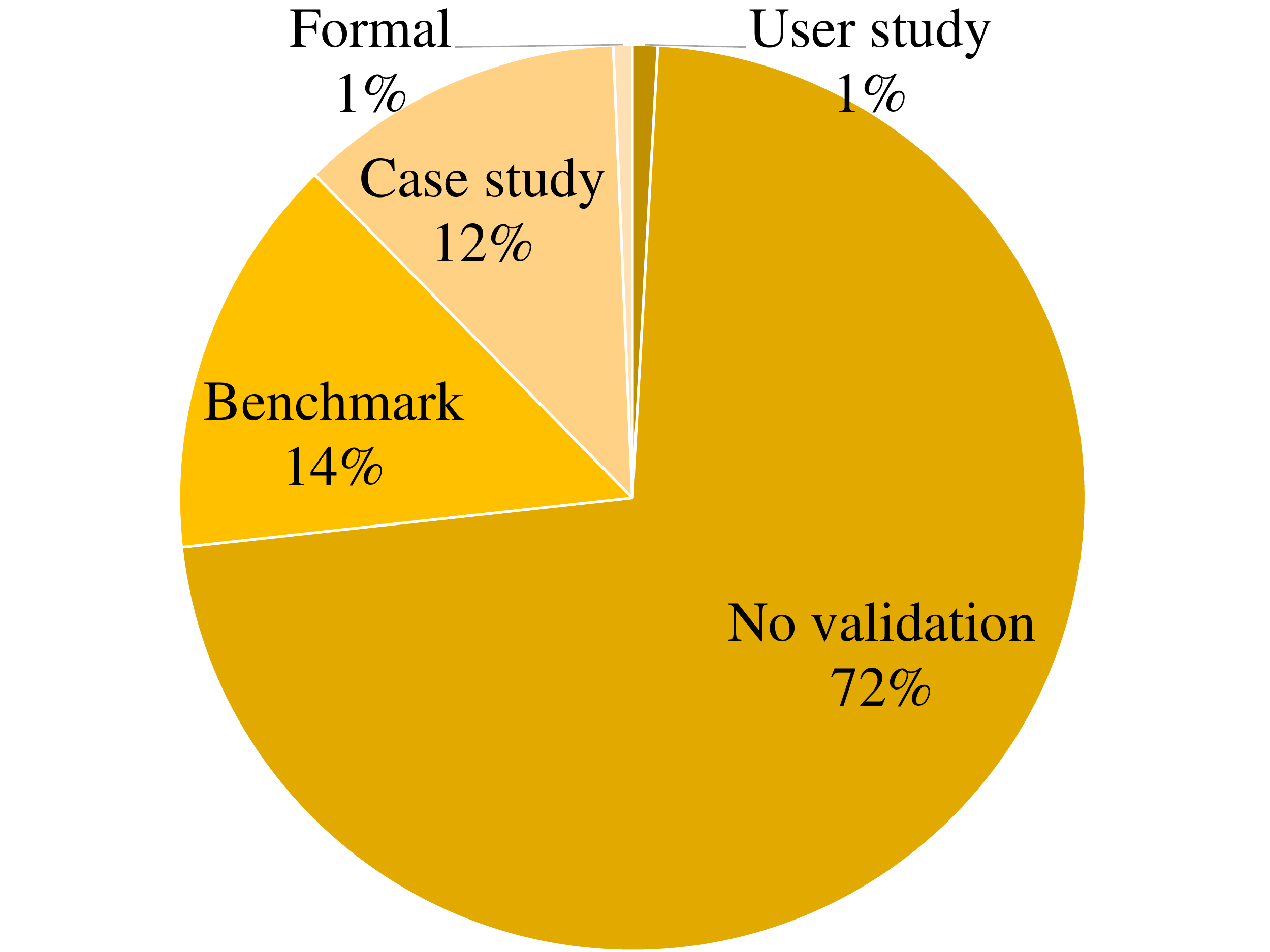}
	\caption{Distribution of validation facet.}
	\label{fig:validation}
\end{figure}
\Fig{fig:validation} represents the distribution of the validation facet a pie chart.
A large majority of the papers did not validate their TBCG approach.
This can be explained in part by the considered papers where TBCG was not the main contribution of the paper.
Furthermore, half of the tools with no validation used were unspecified or rarely used, as in~\cite{Axelsen2011}.
However, some tools have reached a certain maturity and thus dedicate less attention to validation when the tool used offers higher guarantees.
This includes papers such as~\cite{Brox2013} that uses Simulink TLC to generate VHDL code which can be later implemented on reconfigurable Field-Programmable Gate Array devices.
A fair portion of the paper relied on \textbf{benchmark} to validate the TBCG used.
It consists of papers like \cite{Buezas2013} that uses EGL to generate maven code. The authors assess their results using metrics like the time, the gain and the SLOC to validate their work. 
Validation through a \textbf{case study} obtained a similar score.
For example in~\cite{Li2012a}, the authors propose a service-oriented framework with a set of ontology systems to support service and device publishing, discovery and recovery for smart homes. In order to validate their approach, they implement their concept in the particular case of the MediaControl view only.
\textbf{User study} and \textbf{formal} validation account for only eight papers in total.
This is not a surprise since both techniques require more time and resources to setup, unlike the others.
These validation techniques are respectively used in~\cite{Possatto2015} in which eight subjects divided into two groups based on their knowledge and experience with the related technologies performed a series of four tasks and \cite{Fischer2003} where the authors validate their results with the help of mathematical formulas like Gaussian functions.

Moreover, the evolution of no validation class follows the general trend of papers.
We notice that formal validations were used only until 2003.

\subsection{Context}

\begin{figure}[h]
	\centering
	\includegraphics[width=.7\linewidth,trim={0 0 0 14cm},clip]{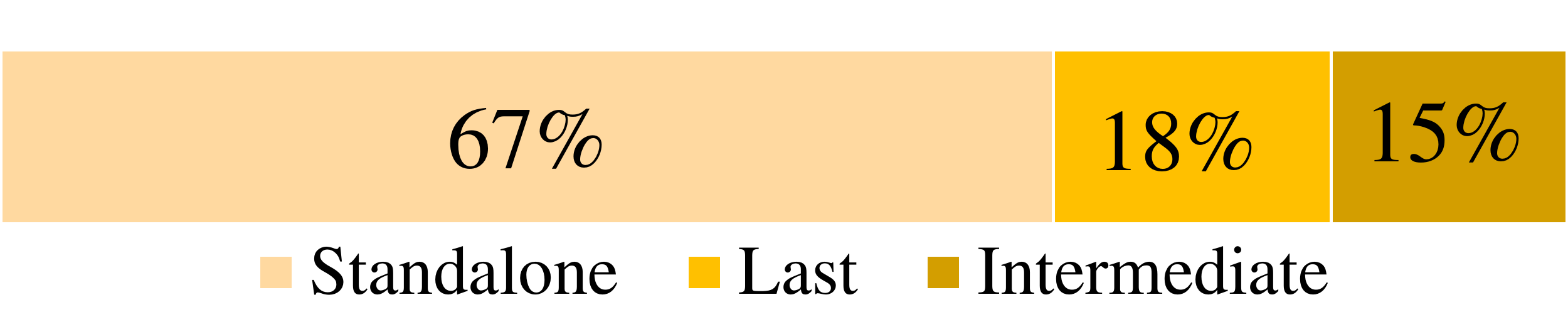}
	\caption{Distribution of context facet.}
	\label{fig:context}
\end{figure}
The distribution of context facet is presented in \Fig{fig:context}.
TBCG was used most of the time \textbf{standalone}, such as in~\cite{Valderas2006}.
The other two classes \textbf{last} and \textbf{intermediate} obtain respectively 18\% and 15\% of the papers.
As an example, TBCG is an intermediate step in~\cite{Sridhara2011} where the generated algorithm is given as one of the inputs of an extraction task. TBCG is the last step of a process in~\cite{Fu2006} that starts with the execution of the various tasks of an integration system and ends with the generation of the final source code.
Most papers only focus on the code generation part but this may have been a part of a bigger project.

\subsection{Orientation}

A quarter (26\%) of the papers in the corpus are (co-)authored by a researcher from \textbf{industry}.
The remaining 74\% are written only by \textbf{academics}.
This is a typical distribution since industrials tend to not publish their work.
This result shows that TBCG is used in industry as in~\cite{Jugel2011}.
Industry oriented papers have gradually increased since 2003 until they reached a peak in 2013.

\subsection{Application domain}

\begin{figure}[h]
	\centering
	\includegraphics[width=\linewidth,trim={0 0 0 8cm},clip]{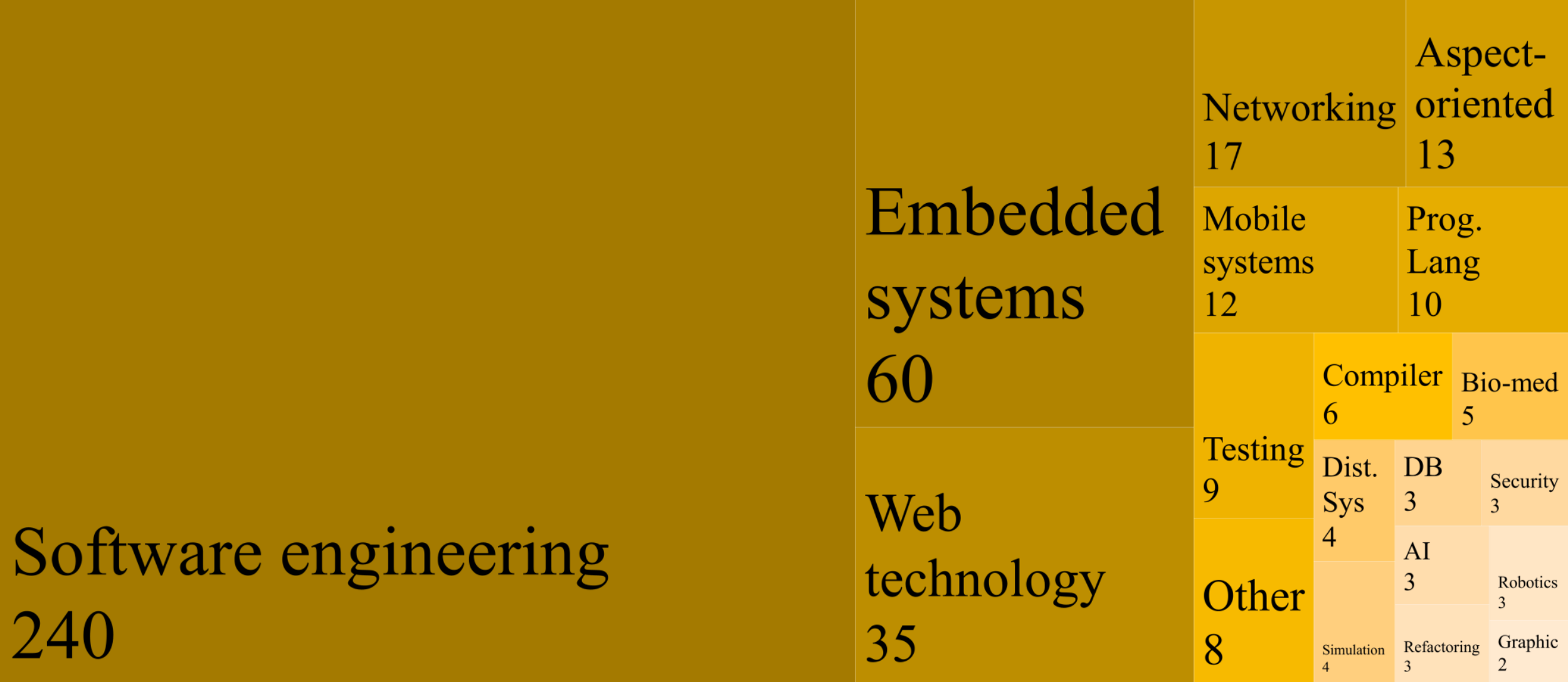}
	\caption{Distribution of application domain facet}
	\label{fig:appdomain}
\end{figure}
The distribution of the application domain facet is shown as a tree map in  \Fig{fig:appdomain}.
It highlights the fact that TBCG is used in many different areas.
\textbf{Software engineering} obtains more than half of the papers with 55\% of the publications.
We have grouped in this category other related areas like ontologies, information systems or software product lines.
This is expected given that the goal of TBCG is to synthesize software applications.
For example, the work in~\cite{Behrens2000} uses the Rational CASE tool to generate Java programs in order to implement an approach that transforms UML state machine to behavioral code.
The next category is \textbf{embedded systems} which obtains 13\% of papers.
Embedded systems often require low level hardware code difficult to write.
Some even consider code generation to VHDL as a compilation rather than automatic programming.
In this category, we found papers like~\cite{Durand2012} in which Velocity is used to produce Verilog code to increase the speed of simulation.
\textbf{Web technology} related application domains account for 8\% of the papers.
It consists of papers like \cite{Schattkowsky2002} where the authors worked to enhance the development dynamic web sites.
\textbf{Networking} obtains 4\% of the papers, such as~\cite{Buezas2013} where code is generated for a telephony service network.
\textbf{Compiler} obtains 1\% of the papers, such as~\cite{Manley2010} where a C code is generated and optimized for an Intel C compiler.
It is interesting to note that several papers were applied in domains such as \textbf{bio-medicine}~\cite{Phillips2006}, \textbf{artificial intelligence}~\cite{Fu2006}, and \textbf{graphics}~\cite{Possatto2015}.
We combined application domains with a single paper into the \textbf{other} category.
This regroups domains such as agronomy, education, and finance.
It is important to mention that the domain discussed in this category corresponds to the domain of application of TBCG employed, which differs from the publication venue.

%% file: correlations.tex
\section{Relations between Characteristics}\label{sec:correlations}

To further characterize the trends observed in \Sect{sec:characteristics}, we identified significant and interesting relations between the different facets of the classification scheme.

\subsection{Statistical correlations} 

%
A Shapiro-Wilk test of each category determined that the none of them are normally distributed.
Therefore, we opted for the Spearman two-tailed test of non-parametric correlations with a significance value of $0.05$ to identify correlations between the trends of each category.
The only significantly strong correlations we found statistically are between the two input types, and between MDE and input type.

 
%

With no surprise, the correlation between \textbf{run-time and design time input} is the strongest among all, with a correlation coefficient of $0.944$ and a \textit{p-}value of less than $0.001$.
This concurs with the results found in \Sect{sec:designtime}.
An example is when the design-time input is UML, the run-time input is always a UML diagram as in~\cite{Phillips2006}. Such a strong relationship is also noticeable in~\cite{Ghodrat2008} with programming languages and source code, as well as in~\cite{Fu2006} when a schema design is used for structured. As a result, all run-time input categories are correlated to the same categories as for design-time input. We will therefore treat these two facets together as \emph{input type}.

There is a strong correlation of coefficient of $0.738$ and a p-value of less than $0.001$ between \textbf{input type and MDE}.
As expected, more than 90\% of the papers using general purpose and domain specific inputs are follow the MDE approach.

\subsection{Other interesting relations}

We also found weak but statistically significant correlations between the remaining facets. We discuss the result here.

\subsubsection{Template style} 

\begin{figure}[h]
	\centering
	\includegraphics[width=\linewidth,trim={0 0 0 1.8cm},clip]{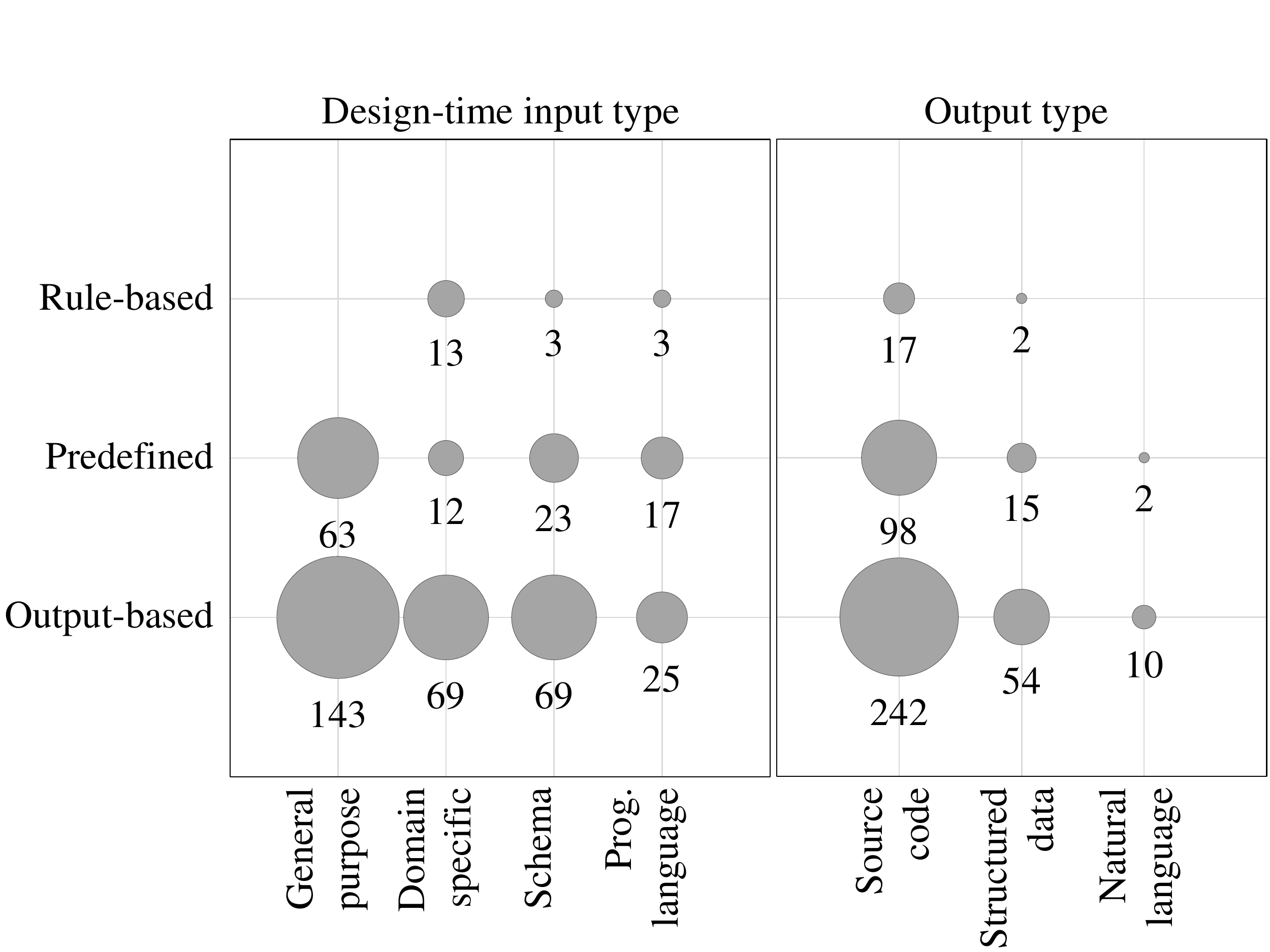}
	\caption{Relation between template style (vertical) and input/output types (horizontal)}
	\label{fig:bubbletemplate}
\end{figure}
\Fig{fig:bubbletemplate} shows the relationship between template style, design-time input, and output types.
We found that for the predefined templates, there are twice as many papers that use schema input than domain specific.
However, for output-based, domain specific inputs are used slightly more often.
We also notice that general purpose input is never used with rule-based templates.
The output type follows the same general distribution regardless of the template style.

We found no rule-based style approach that has validated the TBCG component in their paper.
User studies and formal validations were only performed on approaches using output-based templates.

All rule-based style approaches have included a sample application.
Meanwhile, the proportion of small scale was twice more important for predefined templates (51\%) then for output-based (27\%).

We found that popular tools were used twice more often on output-based templates (58\%) than on predefined templates (23\%).
Rule-based templates never employed a tool that satisfied our popularity threshold, but used other tools such as Stratego.

We found that all papers using a rule-based style template do not follow an MDE approach. On the contrary, 70\% of the output-based style papers and 56\% of the predefined ones follow an MDE approach.

We noted that regardless of the template style, TBCG is used in an intermediate step or at the last step equally often.

Finally, we found that for each template style, the number of papers authored by an industry researcher fluctuated between 22--30\%.

\subsubsection{Input type}

\begin{figure}[h]
	\centering
	\includegraphics[width=.8\linewidth,trim={0 0 7cm 3.8cm},clip]{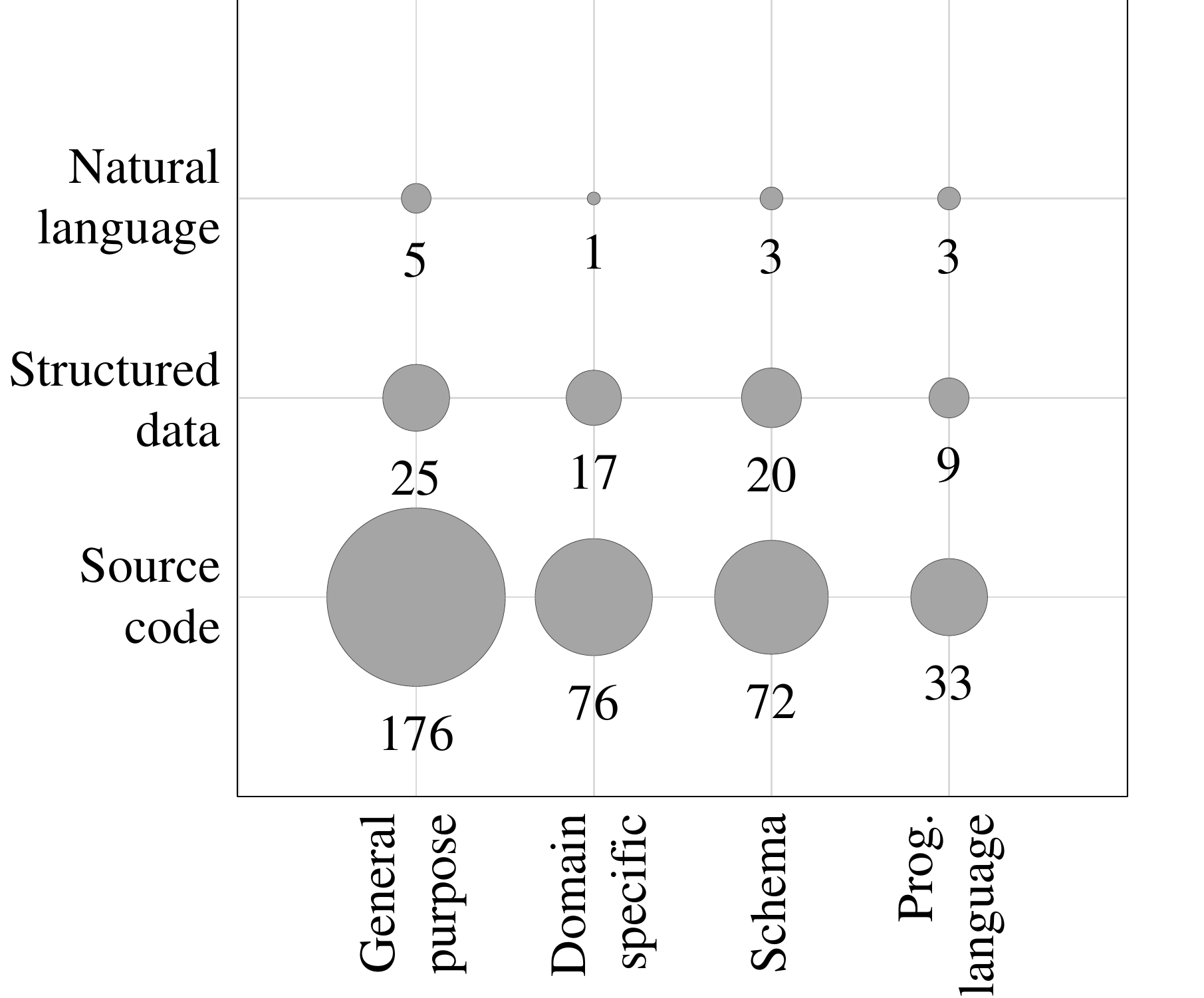}
	\caption{Relation between output (vertical) and design-time input (horizontal) types showing the number of papers in each intersection}
	\label{fig:bubble-input}
\end{figure}

The bubble chart in \Fig{fig:bubble-input} illustrates the tendencies between input and output types.
It is clear that source code is the dominant generated artifact regardless of the input type.
Source code is more often generated from general purpose and domain specific inputs than from schema and programming languages.
Also, the largest portion of structured data is generated from a schema input.
Finally, the most generated natural language text is when source code is provided as input.

Moving on to input type and application scale, we found that small scales are used 40\% of the time when the input is a programming language.
The number of papers with no sample application is very low (5\%) regardless of the template style.
Finally, 74\% of papers using large scale applications use a domain specific input, which is slightly higher than those using a general purpose input with 71\%.

Next, when we compared input type to validation, we found that no paper using a DSL or a programming language used any formal method of validation.
22\% of the papers using a DSL as input used a benchmark to validate their approach, which is higher than the 19\% of the papers using general purpose languages. Also, we found that 77\% of the papers using a general purpose language as input did not validate their approach.

\subsubsection{Output type}

As we compared output type to orientation, we found that industrials generate slightly more source code than academics: 89\% vs. 80\%.
However, academics generate more structured data and natural language than industrials: 18\% vs. 6\% and 3\% vs. 1\% respectively.

\subsubsection{Application scale}
We found that 65\% of the papers without application are from the academy.
Between application scale and tools, we found that 74\% of the papers that make use of a popular tool used large scale application to illustrate their approach.
Also, 62\% of the papers using unpopular tools\footnote{Refers to the union of other and unspecified categories of the tool facet.}
use large scale applications. Small scale is likely to be used in unpopular tools rather than popular tools.

\subsubsection{Validation}

\begin{figure}[h]
	\centering
	\includegraphics[width=.8\linewidth,trim={0 0 4.8cm 1cm},clip]{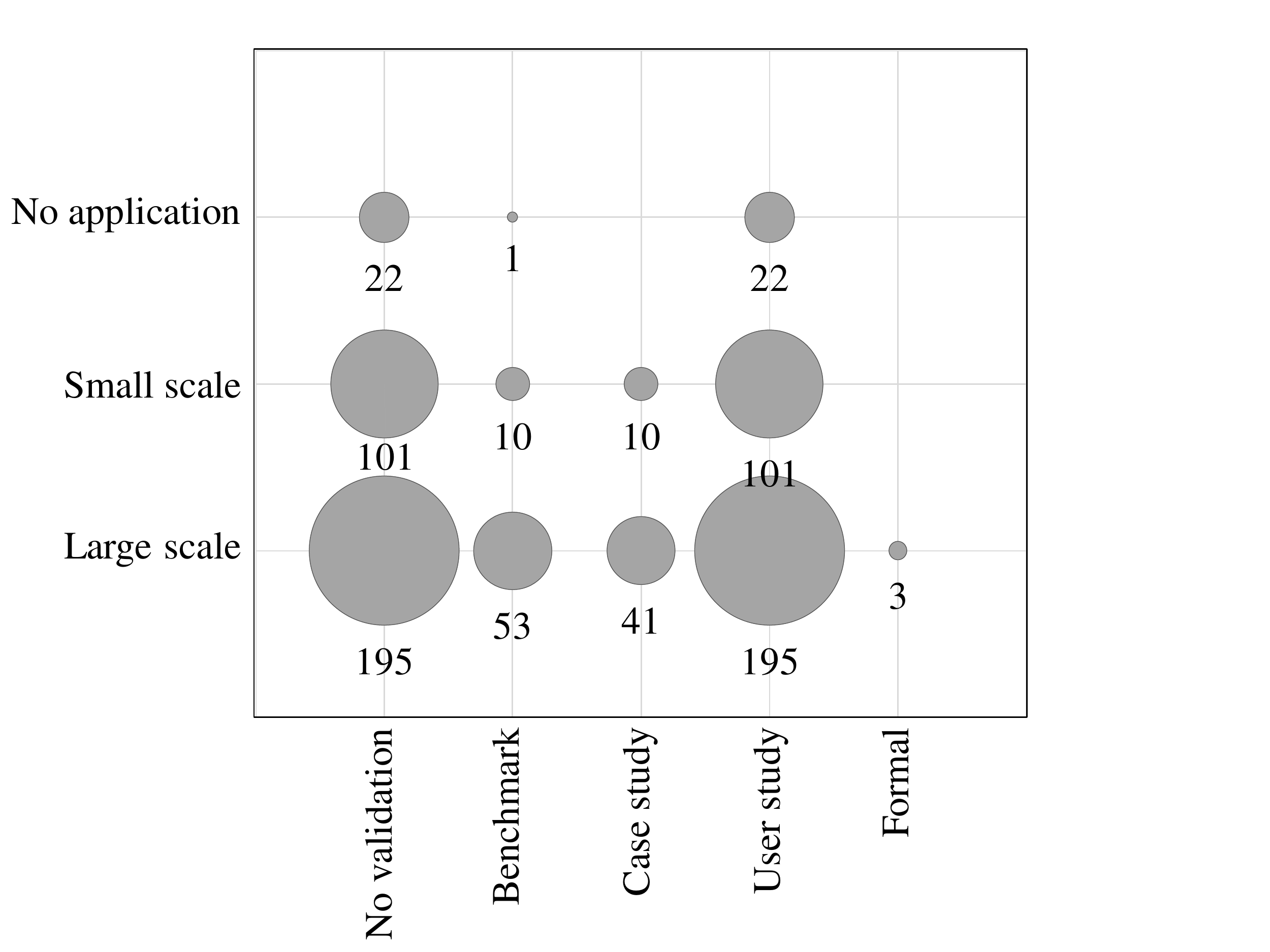}
	\caption{Relation between validation (vertical) and application scale (horizontal)}
	\label{fig:bubble-validation}
\end{figure}

We found that whenever the TBCG is validated, it is always accompanied by application example, as depicted in \Fig{fig:bubble-validation}.
However, when no validation is provided, the paper is still accompanied by an example 67\% of the time.
Large scale applications are used on all instances of user studies and formal validation. 

Between validation and MDE, we noted that 53\% of the papers using benchmark are non MDE. So are all papers with a formal method of validation.
Papers relying on a user study are equally distributed between MDE and non MDE approaches. Furthermore, 68\% of the papers without validation are MDE-based publications.
We found no formal method for validation when using a popular tool. 70\% of the papers using unpopular tools do not validate their approach. 13\% of the papers that use a popular tool conducted a case study. For unpopular tools, a benchmark is used 18\% of the time.

%% file: tools.tex
\section{Template-based Code Generation Tools} \label{sec:tools}

\begin{figure}[h]
	\centering
	\includegraphics[width=0.5\linewidth,trim={0 0 6cm .5cm},clip]{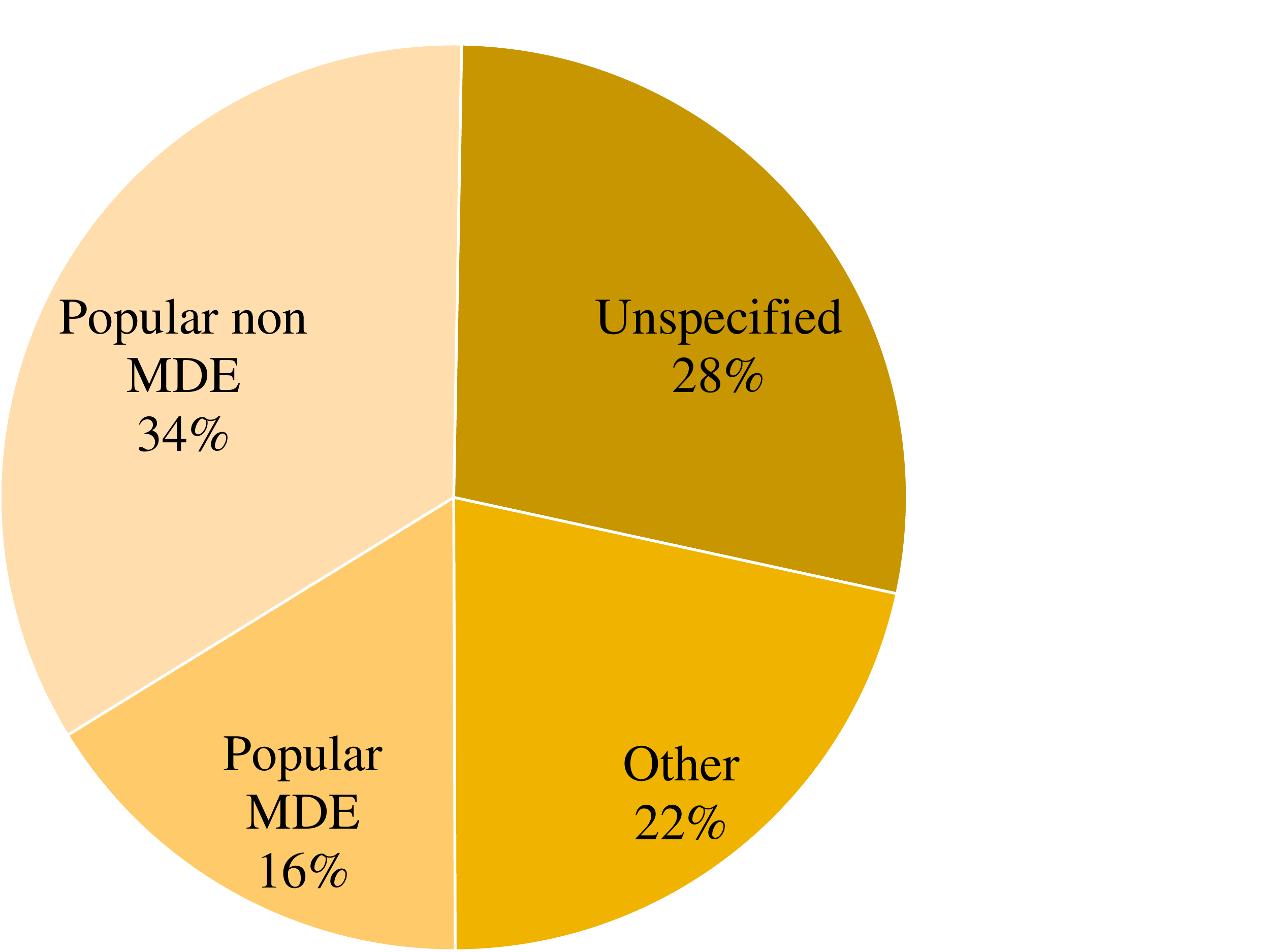}
	\caption{Tools categories}
	\label{fig:tools_categories}
\end{figure}
\Fig{fig:tools_categories} shows that half of the papers used a popular TBCG tool, whereas the other half used less popular tools (the other category), did not mention any TBCG tool, or implemented the code generation directly for the purpose of the paper.
We also see that more than half of the popular tools do not follow MDE approaches.

\subsection{Popular tools}

\begin{figure}[h]
	\centering
	\includegraphics[width=\linewidth,trim={0 0 0 7.5cm},clip]{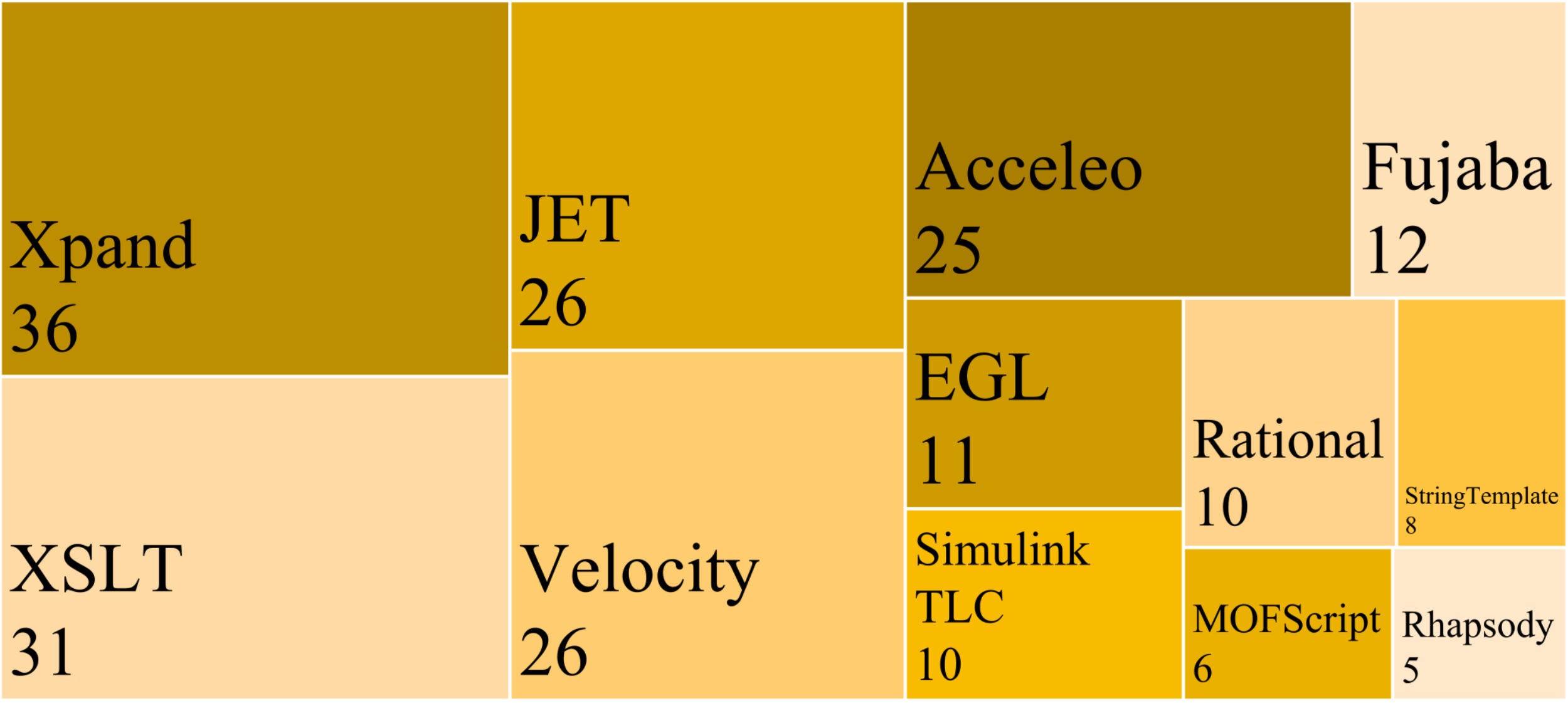}
	\caption{Popular tools}
	\label{fig:namedtools}
\end{figure}
\Fig{fig:namedtools} shows the distribution of popular tools used in at least 1\% of the papers, \ie five papers. Xpand is the most popular with 17\% of the papers using a popular tool. Its popularity is probably due to its simple syntax and ease of use~\cite{Jugel2011}. It is an MDE tool that relies on a metamodel specified in Ecore as design-time input. Xpand templates can be extended with a custom language Xtend\footnote{\url{http://www.eclipse.org/Xtext/documentation/}} or Java, in order to handle more complex tasks.
Acceleo is very similar to Xpand, on top of also being an MDE tool~\cite{Guduvan2013}. They mainly differ in their syntax and that Acceleo implements the M2T specifications standard from the Object Management Group\footnote{\url{http://www.omg.org/spec/MOFM2T/1.0/PDF}}. Acceleo templates can be extended with Java code. Unlike Xpand, the call to the extended class is done directly within the template. Furthermore, Xpand and Acceleo both have an OCL-like language for the dynamic part.
EGL also has a structure similar to the other MDE-based tools. It is natively integrated with languages from the Epsilon family, thus relies on the Epsilon Object Language unlike OCL for the above mentioned tools.
MOFScript is the least used popular MDE-based tool that only differs in syntax from the others.

XSLT is the second most popular tool used.
It is suitable for XML documents only.
Some use it for models represented in their XMI format, as it is the case in~\cite{Adamko2005}.
XSLT follows the template and filtering strategy. It matches each tag of the input document and applies the corresponding template. 

JET~\cite{Kovi2007} and Velocity~\cite{Durand2012} are used as often as each other on top of being quite similar. The main difference is that JET uses an underlying programming language (Java) for the dynamic part. In JET, templates are used to help developers generate a Java class that implements the code generation

StringTemplate has its own template structure. Unlike the above-mentioned tools, to use StringTemplate the developer must write a Java code where strings to be output are defined using templates~\cite{Anjorin2013}. Also, StringTemplate does not allow assignments. Note that all the tools mentioned above use an output-based template style.

The most popular CASE tools for TBCG are Fujaba~\cite{Burmester2005}, Rational~\cite{Brown2005}, and Rhapsody~\cite{Basu2005}. One of the features they offer is to generate different target languages from individual UML elements. All CASE tools (even counting the other category) have been used in a total of 38 papers, which puts them at par with Xpand.
CASE tools are mostly popular for design activities; code generation is only one of their many features.
CASE tools have a predefined template style.

Simulink TLC is the only rule-based tool among the most popular ones. As a rule-based approach, it has a different structure compared to the above mentioned tools. Its main difference is that the developer writes the directives to be followed by Simulink in order to render the final C code from S-functions.

We notice that the most popular tools are evenly distributed between MDE-based tools (Acceleo, Xpand) and non MDE-based tools (JET, XSLT). Surprisingly, XSLT, which has been around the longest, is less popular than Xpand. This is undoubtably explained by the advantages that MDE offers as discussed in \Sect{sec:codeg}.

\subsection{Unspecified and other tools}
As depicted in \Fig{fig:tools_categories}, 28\% of the papers did not specify the tool that was used, as in~\cite{Furusawa2010} where the authors introduce the concept of a meta-framework to resolve issues involved in extending the life of applications. 
Furthermore, 22\% of the papers used less popular tools, present in less than five papers, such as T4 and Cheetah used in~\cite{Manley2010}, which is a python powered template mainly use used for web developing. Cheetah templates are class definition. The generated classes can either be used immediately or given into a python module. Some CASE tools were also in this category, such as AndroMDA~\cite{Muller2005}. Other examples of less popular tools are FreeMarker~\cite{Munoz2006}, Meta-Aspect-J~\cite{Antkiewicz2006}, and Zerberus~\cite{Fischer2003}.
The fact that new or less popular tools are still abundantly used suggests that research in TBCG is still active with new tools being developed or evolved.

\subsection{Trends of tools used}
Each one of these tools had a different evolution over the years. Unspecified tools were prevailing before 2004 and then kept a constant rate of usage until a drop since 2014. We notice a similar trend for CASE tools that were the most popular in 2005 before decreasing until 2009. They only appear in at most three papers per year after 2010. The use of the most popular tool, Xpand, gradually increased since 2005 to reach the peak in 2013 before decreasing. Other category maintained an increasing trend until 2014. Yet, a few other popular tools appeared later on. For example, EGL started appearing in 2008 and had its peak in 2013. Acceleo appeared a year later and was the most popular TBCG tool in 2013--2014. Finally, MOFScript had no more than a paper per year since 2005. StringTemplate and T4 were used scarcely respectively since 2006 and 2009.

\subsection{Characteristics of tools} 
We have also analyzed each popular tool with respect to the characteristics presented in \Sect{sec:characteristics}. As mentioned earlier, most of the popular tools implement output-based template technique except the CASE tools which are designed following the predefined style.

Tools such as Acceleo, Xpand, EGL, MOFScript and 97\% of the CASE tools papers are only used based on an MDE approach, given that they were created by this community. Nevertheless, there are tools that were never used with MDE principles, like T4. Such tools can handle a program code or a schema as metamodel but have no internal support for modeling languages. Moreover, the programmer has to write his own stream reader to parse the input, but they allow for a broader range of artifacts as inputs that do not have to be modeled explicitly.
Between MDE-based and non MDE-based tools, we have few that provide internal support for both MDE-based and non MDE-based approaches. In fact, tools like Velocity, XSLT and StringTemplate can handle both UML metamodels and programmed metamodel.

A surprising result we found is that EGL is the only MDE tool that has its papers mostly published in MDE venues like \textsc{Sosym}, \textsc{Models}, and \textsc{Ecmfa}. All the other tools are mostly published in other venues like \textsc{Icssa}, whereas software engineering venues, like \textsc{Ase} or \textsc{Icse}, and MDE venues account for 26--33\% of the papers for each of the rest of the MDE tools.

CASE tools, MOFScript, Velocity, and Simulink TLC mostly generate program code. The latter is always used in the domain of embedded systems. Papers that use StringTemplate do not include any validation process, so is Velocity in 93\% of the papers using it. XSLT has been only used to generate structured data as anticipated.

Other tools are the most used TBCG in the industry. This is because the tool is often internal to the company~\cite{Kulkarni2011}. Among the most popular tools, Xpand is the most in the industry.

%% file: mde.tex
\section{MDE and Template-based Code Generation} \label{sec:mde}

Overall, 64\% of the publications followed MDE techniques and principles. For example in paper \cite{Touraille2011}, the authors propose a simulation environment with an architecture that aims at integrating tools for modeling, simulation, analysis, and collaboration. As expected, most of the publications using output-based and predefined techniques are classified as MDE-based papers.
The remaining 36\% of the publications did not use MDE. This includes all papers that use a rule-based template style as reported in \Sect{sec:correlations}. For example, the authors in~\cite{Buckl2005} developed a system that handles the implementation of dependable applications and offers a better certification process for the fault-tolerance mechanisms.

\begin{figure}[h]
	\centering
	\includegraphics[width=.7\linewidth]{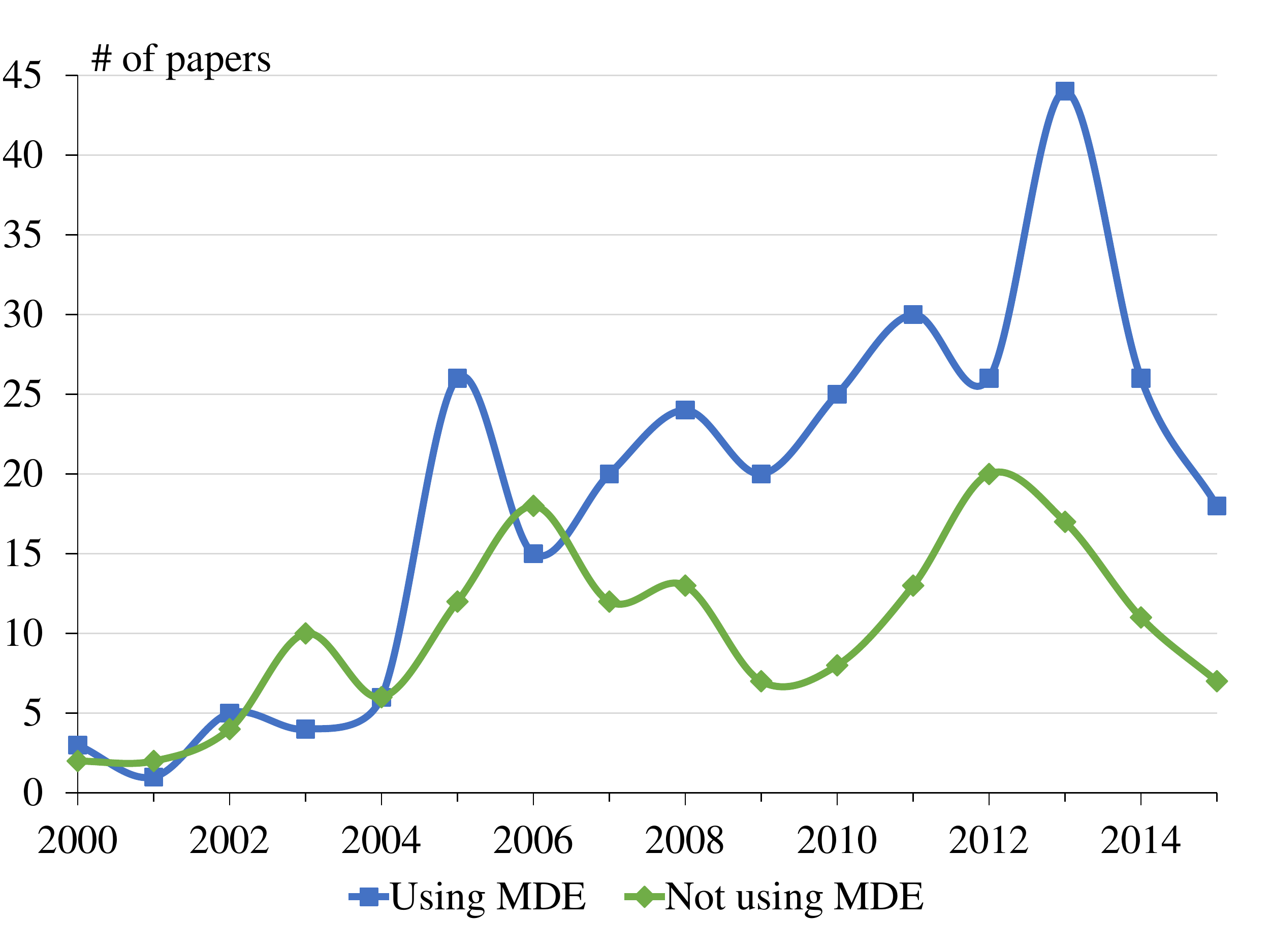}
	\caption{Evolution of the MDE facet}
	\label{fig:mdetrends}
\end{figure}
As \Fig{fig:mdetrends} shows, the evolution of the MDE category shows that MDE-based approach started overpassing non MDE-based techniques in 2005, with the exception of 2006. It increased to reach a peak in 2013 and then started decreasing as the general trend of the corpus. Overall, MDE-based technique for TBCG has been dominating other techniques in the past 10 years. 

We also analyzed the classification of only MDE papers with respect to the characteristics presented in \Sect{sec:methods}. We only focus here on facets with different results compared to the general trend of papers.
We found that only half of the total number of papers using unspecified and other tools are MDE-based papers.
We only found one paper that uses a programming language as design-time input with MDE~\cite{Fertalj2002}. This analysis also shows that the year 2005 clearly marked the shift from schema to domain-specific design-time inputs, as witnessed in \Sect{sec:designtime}. Thus after general purpose, which obtains 69\% of the publications, domain specific accounts a better score of 26\%, while schema obtains only 4\%.
With respect to the run-time category, the use of domain-specific models increased to reach a peak in 2013. As expected, no program code is used for MDE papers, because MDE typically does not consider them as models, unless a metamodel of the programming language is used.
We notice that when we focus only on MDE publications, case studies are more used to validate the TBCG than benchmarks are.

Interestingly, MDE venues are only the second most popular after other venues for MDE approaches.
Finally, MDE journal papers maintained a linear increase over the years, while MDE conference papers had a heterogeneous evolution similar to the general trend of papers.

%% file: discussion.tex
\section{Discussion} \label{sec:discussion}

\subsection{RQ1: What are the trends in TBCG?} \label{sec:rq1}

Following a deep investigation of the statistical results from the classification, we noticed that TBCG has received sufficient attention from the research community.
Even though the number of papers has been decreasing for the past two years, the community has maintained a production rate in-line with the last ten years average, especially with a constant rate of appearance in journal articles.
This brings us to the same conclusion as Batot \etal{}~\cite{Batot2016}, that TBCG has reached a certain maturity from a research point of view in 2013.
The lack of retention of papers appearing in non MDE may indicate that TBCG is now applied in development projects rather than being a critical research problem to solve.
As it is common in other domains, the production of TBCG papers has also been influenced by major events in 2005 and 2013.
MDE has attracted more attention to the code synthesis technique.
Also, conference papers as well as non MDE and software engineering venues had a significant impact on the evolution of TBCG.
Finally, TBCG seems to have reached a steady publication rate since 2005.
Hence, we can expect contributions from the research community to continue in that trend. 

\subsection{RQ2: What are the characteristics of TBCG approaches?} \label{sec:rq2}

Our classification scheme constitutes the main source to answer this question. The results clearly indicate the preferences the research community has regarding TBCG. Output-based template styles have always been the most popular from the beginning. Nevertheless, there have been some attempts to propose other template styles like the rule-based style or the predefined style but they did not catch on. Because of its simplicity to use, the predefined style is probably still popular in practice, but it is less mentioned in research papers. TBCG has been used to synthesize a variety of application code or documents. As expected, the study shows that high-level language inputs have prevailed over any other type. Specifically for MDE approaches to TBCG, the input to transform is moving from general purpose to domain-specific models. Also, the study shows that TBCG is mostly used as a standalone process, unlike what was predicted in MDE~\cite{Kleppe2003}. This indicates that it is yet to be integrated in the development process. It is more like a distinct task that is used only when needed. Academic researchers have contributed most, as expected with a literature review, but we found that industry is actively and continuously using TBCG as well. The study also shows that the community is moving from large scale applications to smaller sized examples in research papers. This concurs with the level of maturity of this synthesis approach as discussed in \Sect{sec:rq1}. 
The study confirms that the community uses TBCG to generate mainly source code. This trend is set to continue since the automation of computerized tasks is continuing to gain ground in all fields. Finally, TBCG has been implemented in many domains. Software engineering and embedded systems are the most popular, but it is also used unexpectedly in unrelated domains like bio-medicine and finance.

\subsection{RQ3: To what extent are TBCG tools being used?} \label{sec:rq3}

In this study, we discovered a total 70 different tools for TBCG. Many studies implemented code generation with a custom-made tool that was never or seldom reused. This indicates that the development of new tools is still very active. MDE tools are the most popular. Since the research community has favored output-based template style (c.f. \Sect{sec:rq2}), this has particularly influenced the tools implementation. 
This template style allows for more fine-grained customization of the synthesis logic which seems to be what users have favored. This particular aspect is also influencing the expansion of TBCG into industry. Well-known tools like Acceleo, Xpand and Velocity are moving from being simple research material to effective development resources in industry. Finally, the study shows that there is has been a shift from CASE tools to output-based tools since 2005.

\subsection{RQ4: What is the place of MDE in TBCG?} \label{sec:rq4}

All this analysis clearly conclude that the advent of MDE has gathered high attention to TBCG. In fact, MDE has lead to increase the average number of publications by a factor of four. 
As TBCG became a commonplace in general, the research in this area is now mostly conducted by the MDE community. Also, MDE has brought very popular tools that have encountered a great success and they are also contributing to the expansion of TBCG across industry. It is important to mention that the MDE community publishes in specific venues like \textsc{Models}, \textsc{Sosym}, or \textsc{Ecmfa} unlike other research communities. This resulted in three MDE venues at top of the ranking (c.f. \Table{tab:venues}).

\subsection{Threats to validity} \label{threats}

The results presented in this systematic mapping study have depended on many factors that could potentially limit the study.

\subsubsection{Construction validity}

Threats to construction validity deals with the problems related to the design of the research method and especially to identifying relevant primary studies.

In a strict sense, our findings are valid only for our sample that we collected from 2000--2015. This leads to determine whether the primary studies used in our SMS are good representation of the whole population. From \Fig{fig:general_trend}, we can observe that our sample can be attributed as a representative sample of the whole population. In particular, the average number of identified primary studies per year is 27.8 with standard deviation 15.75. Since it is difficult to be exhaustive on TBCG, we selected three of the major online databases. These databases are complementary and we are confident that they index a maximum of relevant publications. We chose to obtain the best possible coverage at the cost of duplications. 

Another potential limitation is the search query. It is difficult to encode a query that is restrictive enough to discard unrelated publications but at the same time retrieves all the relevant ones. In order to obtain a satisfactory balance, we included synonyms and captured possible declinations. Our search query could suggest a restriction of the type of output. However, the size of the final corpus we classified is about ten times larger than other SMS related to code generation (see \Sect{sec:sms}).
We are therefore confident that the final corpus is a representative subset of all relevant publications on TBCG. 

Finally, given that we obtained a sufficiently large final corpus for typical SMS, we did not perform snowballing which may have resulted in collecting additional papers omitted by the search engines.

\subsubsection{Internal validity}

A potential limitation is related to data extraction. It is difficult to extract data from relevant publications especially when the quality of the paper is low, when code generation is not the primary contribution of the paper, or when critical information for the classification is not directly available in the paper. For example in \cite{Ma2012}, the authors only mention the name of the tool used to generate the code. In order to mitigate this threat, we had to resort to searching for additional information about the tool: reading other publications that use the tool, traversing the website of the tool, installing the tool, or discussing with the tools experts, as reported in \Sect{sec:eligibility}.

Another possible threat is the screening of papers based on inclusion and exclusion criteria that we defined before the study was conducted. During this process, we examined only the title, the abstract. Therefore, there is a probability that we excluded relevant publications such as \cite{Buezas2013}, that do not include any TBCG terms. In order to mitigate this threat, whenever we were unsure whether a publication should be excluded or not we conservatively opted to include it. However, during classification when reading the whole content of the paper, we may still have excluded it. 

\subsubsection{External validity}

External threats to validity cope with problems that might arise during conclusion generalization.
The results we obtained are based on TBCG only. Even though our classification scheme includes facets like validation, orientation, application domain, that are not related to the area, we followed a topic based classification. The core characteristics of our study are strictly related to this particular code synthesis technique. We have defined characteristics like template style and the two levels of inputs that we believe are exclusive to TBCG. Therefore, the results cannot be generalized to other code generation techniques mentioned in \Sect{sec:cg-techniques}.

\subsubsection{Conclusion validity}

Threats to conclusion validity (or reliability) deal with problems that might arise when deriving conclusions and whether the SMS can be repeated.
Our study is based on a large number of primary studies. This helps us mitigate the potential threats related to the conclusions of our study. A missing paper or a wrongly classified paper would have a very low impact on the statistics compared to a smaller number of primary studies. In addition, as a senior reviewer did a sanity check on the rejected papers, we are confident that we did not miss a significant number of papers. Hence, the chances for wrong conclusions are small.
Replication of this study can be achieved as we provided  all the details of our research method in \Sect{sec:methods}. Also, our study follows the methodology described in \cite{Petersen2008}.

%% file: conclusion.tex
\section{Conclusion} \label{sec:conclusion}

This paper reports the results of a SMS we conducted on the topic of TBCG, which has been missing in the literature. The objectives of this study are to better understand the characteristics of TBCG techniques and associated tools, identify research trends, and assess the importance of the role that MDE plays.
We have systematically scanned the published, peer-reviewed literature and studied an extensive set of 440 papers published during the period 2000--2015.
The analysis of this corpus is organized into facets of a novel classification scheme, which is of great value to modeling and software engineering researchers who are interested in painting an overview of the literature on TBCG.

Our study shows that the community has been diversely using TBCG over the past 15 years and that research and development is still very active. TBCG has greatly benefited from MDE in 2005 and 2013 which mark the two peaks of the evolution of this area, tripling the average number of publications. In addition, TBCG has favored a template style that is output-based and high level modeling languages as input. TBCG is mainly used to generate source code and has been applied in a variety of domains. The community has been favoring the use of custom tools for code generation over popular ones. Furthermore, both MDE and non-MDE tools are becoming effective development resources in industry. Finally, we found that there is a lack of a formal verification method for TBCG approaches, the existing ones being targeted to compilers. Nevertheless, this can be a good starting point in the process of formalizing verification methods that suit TBCG. 


As future work, we would like to revise the query to include not only ``code'' as the main output, but all the other possible artifacts such as documents. We would also like to pursue the study for a couple of years past 2015 to justify the observation of the 2013 event. Finally we would like to analyze and compare the tools (thus extending the work in~\cite{Rose2012}) to help users decide when to use which tool.